\documentclass[preprint]{aastex63}
\usepackage{amssymb}
\usepackage{float} 
\shorttitle{The Most Rapidly Rotating Ultra-Cool Dwarfs}
\shortauthors{Tannock et al.}
\usepackage{gensymb}
\usepackage{natbib}
\setcitestyle{notesep={; }}
\usepackage{longtable}
\accepted{to the Astronomical Journal March 1, 2021}

\begin{document}

\title{Weather on Other Worlds. V. The Three Most Rapidly Rotating Ultra-Cool Dwarfs}

\author[0000-0002-9445-2870]{Megan E. Tannock}
\affiliation{Department of Physics and Astronomy, The University of Western Ontario, 1151 Richmond St, London, Ontario, N6A~3K7, Canada}
\correspondingauthor{Megan E. Tannock}
\email{mtannock@uwo.ca}

\author[0000-0003-3050-8203]{Stanimir Metchev}
\affiliation{Department of Physics and Astronomy, Institute for Earth and Space Exploration, The University of Western Ontario, 1151 Richmond St, London, Ontario, N6A~3K7, Canada}
\affiliation{Department of Astrophysics, American Museum of Natural History, 200 Central Park West, New York, New York, 10024--5102, USA}

\author[0000-0003-3313-4921]{Aren Heinze}
\affiliation{Institute for Astronomy, University of Hawaii, 2680 Woodlawn Dr, Honolulu, Hawaii, 96822, USA}

\author[0000-0003-2446-8882]{Paulo A. Miles-P{\'a}ez}
\affiliation{European Southern Observatory, Karl-Schwarzschild-Strasse 2, 85748 Garching, Germany}

\author[0000-0002-2592-9612]{Jonathan Gagn\'e}
\affiliation{Plan\'etarium Rio Tinto Alcan, Espace pour la Vie, 4801 av. Pierre-de Coubertin, Montr\'eal, Qu\'ebec, H1V~3J3, Canada}
\affiliation{Institute for Research on Exoplanets, Universit\'e de Montr\'eal, D\'epartement de Physique, C.P.~6128 Succ. Centre-ville, Montr\'eal, Qu\'ebec, H3C~3J7, Canada}

\author[0000-0002-6523-9536]{Adam Burgasser} 
\affiliation{Department of Physics, University of California, San Diego, 9500 Gilman Dr, La Jolla, California, 92093, USA}

\author[0000-0002-5251-2943]{Mark S. Marley} 
\affiliation{NASA Ames Research Center, Mountain View, California, 94035, USA}

\author[0000-0003-3714-5855]{D\'aniel Apai} 
\affiliation{Steward Observatory, The University of Arizona, 933 N Cherry Ave, Tucson, Arizona, 85721, USA}
\affiliation{Lunar and Planetary Laboratory, The University of Arizona, 1629 E University Blvd, Tucson, Arizona, 85721--0092, USA}

\author[0000-0002-2011-4924]{Genaro Su{\'a}rez}
\affiliation{Department of Physics and Astronomy, The University of Western Ontario, 1151 Richmond St, London, Ontario, N6A~3K7, Canada}

\author[0000-0002-8864-1667]{Peter Plavchan} 
\affiliation{Department of Physics and Astronomy, George Mason University, 4400 University Dr, Fairfax, Virginia, 22030, USA}

\begin{abstract}

We present the discovery of rapid photometric variability in three ultra-cool dwarfs from long-duration monitoring with the Spitzer Space Telescope. The T7, L3.5, and L8 dwarfs have the shortest photometric periods known to date: ${1.080}^{+0.004}_{-0.005}$~h, ${1.14}^{+0.03}_{-0.01}$~h, and ${1.23}^{+0.01}_{-0.01}$~h, respectively. We confirm the rapid rotation through moderate-resolution infrared spectroscopy, which reveals projected rotational velocities between 79 and 104~km~s$^{-1}$. We compare the near-infrared spectra to photospheric models to determine the objects' fundamental parameters and radial velocities. We find that the equatorial rotational velocities for all three objects are $\gtrsim$100~km~s$^{-1}$. The three L and T dwarfs reported here are the most rapidly spinning and likely the most oblate field ultra-cool dwarfs known to date. Correspondingly, all three are excellent candidates for seeking auroral radio emission and net optical/infrared polarization. As of this writing, 78 L-, T-, and Y-dwarf rotation periods have now been measured. The clustering of the shortest rotation periods near 1 h suggests that brown dwarfs are unlikely to spin much faster. 
\end{abstract}   


\section{Introduction} 
\label{Introduction}

Variability in ultra-cool dwarfs (spectral types $>$ M7;  \citealt{Kirkpatrick1997}) is caused by large-scale atmospheric structures, such as spots or longitudinal bands \citep{Artigau2009,Radigan2014a,Apai2017}. As inhomogenities rotate in and out of view, they change the object's observed flux on the time scale of the rotation period \citep{Tinney1999, Bailer-Jones2002}.
The largest and most sensitive ultra-cool dwarf monitoring surveys (e.g.,  \citealt{Radigan2014a, Radigan2014b, Buenzli2014, Metchev2015}) have found that variability is common across L and T dwarfs. 
\citet{Metchev2015} estimate that $53\%^{+16\%}_{-18\%}$ of L3--L9.5 and $36\%^{+26\%}_{-17\%}$ of T0--T8 dwarfs are variable at $> 0.4\%$. The rotational periods inferred for L and T dwarfs range over at least an order of magnitude: from 1.4~h \citep{Clarke2008} to likely longer than 20~h \citep{Metchev2015}. Spectroscopic observations have shown that many ultra-cool dwarfs have relatively large projected rotational velocities ($v\sin i \ge 10$\,km\,s$^{-1}$; \citealt{Mohanty2003,Basri2000,Zapatero2006,Reiners2008,Reiners2010,Blake2010,Konopacky2012}), and in some cases rotate at $\sim$30\,\% of their break-up speed (e.g., \citealt{Konopacky2012}). Ultra-cool dwarfs with halo kinematics also exhibit rapid rotation \citep{Reiners2006}, indicating that they maintain relatively large rotational velocities during their entire lifetimes. 

In this paper we present the discovery of three ultra-cool dwarfs with the shortest known photometric---and likely rotational---periods. In Section~\ref{sec:photometry} we present our photometric monitoring with the Spitzer Space Telescope (Spitzer) and the discovery of the short periodicities.  In Section~\ref{sec:spectra} we present moderate-resolution infrared spectroscopy to confirm the rapid rotation of each target.  In Section~\ref{sec:SpectraAnalysis} we fit photospheric models to the spectra to determine the objects' projected rotational velocities and physical parameters, and find the highest $v\sin i$ value yet reported for ultra-cool dwarfs. We discuss the objects' rapid spins and oblateness in Section~\ref{sec:Discussion}. Our findings are summarized in Section \ref{Conclusions}.

\section{Spitzer Photometry, Variability, and Periods}
\label{sec:photometry}

The photometric observations were obtained as part of the GO 11174 (PI: S.\ Metchev) Spitzer Exploration Science Program, ``A Paradigm Shift in Substellar Classification: Understanding the Apparent Diversity of Substellar Atmospheres through Viewing Geometry.'' The program targeted 25 of the brightest known L3--T8 dwarfs to complement our earlier sample of 44 photometrically monitored L and T dwarfs \citep{Metchev2015} and to investigate viewing geometry effects on photometric variability and brown dwarf colors. A full description of the program will be presented in a later publication.

We focus on three variables from the GO 11174 Spitzer program with photometric periods shorter than the shortest previously known: the 1.41 $\pm 0.01$ h period of 2MASS J22282889$-$4310262 \citep{Clarke2008,Buenzli2012,Metchev2015}. Our targets are: the L3.5 dwarf 2MASS J04070752$+$1546457 (\citealt{Reid2008}; herein 2MASS J0407+1546), the L8 dwarf 2MASS J12195156$+$3128497 (\citealt{Chiu2006}; herein 2MASS J1219+3128), and the T7 dwarf 2MASS J03480772$-$6022270 (\citealt{Burgasser2003}; herein 2MASS J0348$-$6022).

\subsection{Warm Spitzer Observations} 
\label{sec:spitzer_observations}

We observed the three objects in staring mode with Spitzer's Infrared Array Camera's (IRAC; \citealt{Werner2004}, \citealt{Fazio2004}) channels 1 (3.6 $\mu$m, [3.6]) and 2 (4.5 $\mu$m, [4.5]). The dates of the observations are given in Table~\ref{table:fastperiods}. The observing sequence was a 10 h staring observation in channel 1, followed immediately by a 10 h staring observation in channel 2. All exposures were 12 s long, taken in full-array readout mode. At the beginning of each staring sequence an additional 0.5 h were used for pointing calibration with the Pointing Calibration and Reference Sensor (PCRS). The PCRS peak-up procedure is intended to correct telescope pointing over long staring observations. We used nearby bright stars for peak-up, as none of our targets were sufficiently bright to perform the peak-up on-target. 

The Spitzer IRAC detector is subject to intrapixel sensitivity variations, known as the ``pixel phase effect'' \citep{Reach2005}.  Precise photometry requires correcting for an object's positioning to sub-pixel precision.  The pixel phase effect is well characterized in a $0.5\times0.5$ pixel ($0\farcs6\times0\farcs6$) ``sweet spot'' \citep{Mighell2008} near a corner of the IRAC array, and flux correction routines are available at the Spitzer Science Centre IRAC High Precision Photometry website.\footnote[13]{\url{https://irachpp.spitzer.caltech.edu}}

We sought to acquire our targets as closely as possible to the center of the IRAC sweet spot.  We used observation epoch-dependent positional corrections for proper and parallactic motions derived from a 2MASS-AllWISE cross-correlation.  However, we were not entirely successful. The average centroid position for each of our three targets was up to a pixel away from the center of the sweet spot: i.e., twice its half-width. We therefore used our own custom pixel phase correction code \citep{Heinze2013} developed for the Spitzer Cycle 8 ``Weather on Other Worlds'' program \citep{Metchev2015} and summarized in Section~\ref{sec:spitzer_variability}.  Experiments with archival Spitzer GO~13067 data on TRAPPIST-1, which was acquired on the sweet spot, confirmed that for point sources at least as bright as TRAPPIST-1 (WISE $W1=10.07$~mag), our pixel phase correction approach is at least as accurate as the set of sweet spot pixel phase corrections on the IRAC High Precision Photometry website.

\subsection{Photometry and Initial Variability Assessment}
\label{sec:spitzer_variability}

We conducted a two-stage photometric and variability assessment, using the Spitzer Basic Calibrated Data images. We first performed approximate [3.6]- and [4.5]-band photometry in 1.5 pixel-radius apertures with the {\sc IDL} Astronomy User's Library\footnote[14]{\url{https://idlastro.gsfc.nasa.gov}} task {\sc aper}. We applied the corresponding aperture correction from the IRAC Instrument Handbook, and a custom pixel phase correction derived as a two-dimensional quadratic function of the centroid position on the detector.

\begin{figure*}
  \centering 
    \includegraphics[trim={1.9cm 6.8cm 3cm 12.2cm}, clip,width=0.98\textwidth]{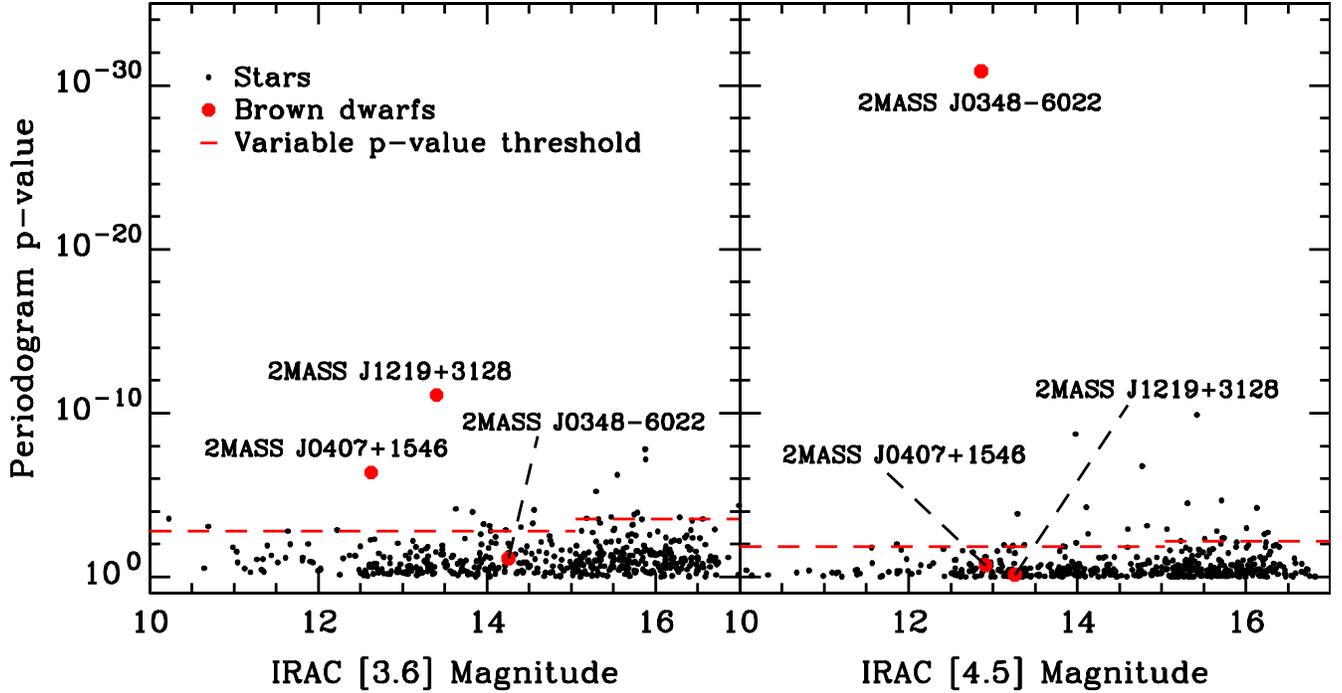}
  \caption{
  Results from the initial periodogram-based variability assessment on the three L and T dwarfs (red points) in the two Spitzer bands, compared to 469 other stars (black points) in the IRAC field of view for our full Spitzer sample. The horizontal dashed lines mark the $p$-value thresholds below which we claim variability, with 95\% of the comparison stars below this line. We separately compute the 95\% threshold for each IRAC channel for the brighter half of comparison stars (log($p$-value) $ = -2.8$ and $-1.9$ in [3.6] and [4.5], respectively) and the fainter half of companion stars (log($p$-value) $ = -3.5$ and $-2.2$ in [3.6] and [4.5], respectively).
  }
  \label{fig:FAPvsmag}
\end{figure*}

We identified variable targets by Lomb-Scargle periodogram analysis \citep{Scargle1982}, sampling periods between 0.1 h and the full 20 h duration of our Spitzer observations. We use the $p$-value, a measure of the likelihood that any variations are caused by random noise, to determine the significance of the periodogram peaks. The $p$-value is $Me^{-P}$, where $P$ is the periodogram power of the highest peak, and $M$ is the number of independent periods considered \citep{Scargle1982, Press1992}. Relative to other non-variable field stars, a sinusoidal signal yields the lowest $p$-value at a given amplitude, making it ideal for detecting rotation-induced photometric variations. We determined a threshold to identify variables by calculating the $p$-value from pixel phase-corrected light curves of 469 field stars in the IRAC field of view for our full Spitzer sample, with obvious variables (e.g., eclipsing binaries) rejected by visual inspection. We split the field stars into two equal-sized groups based on their magnitudes. From the full Spitzer sample, we selected as candidate variables those L and T dwarfs for which the $p$-value was below that of 95\% of field star $p$-values. The $p$-values of the current three L and T dwarfs and of the field stars are shown in Figure~\ref{fig:FAPvsmag}. The relevant thresholds are shown as dashed horizontal lines. 

\begin{figure*}
  \centering
    \includegraphics[trim={0.8cm 9.3cm 0.8cm 4.1cm},clip,width=0.99\textwidth]{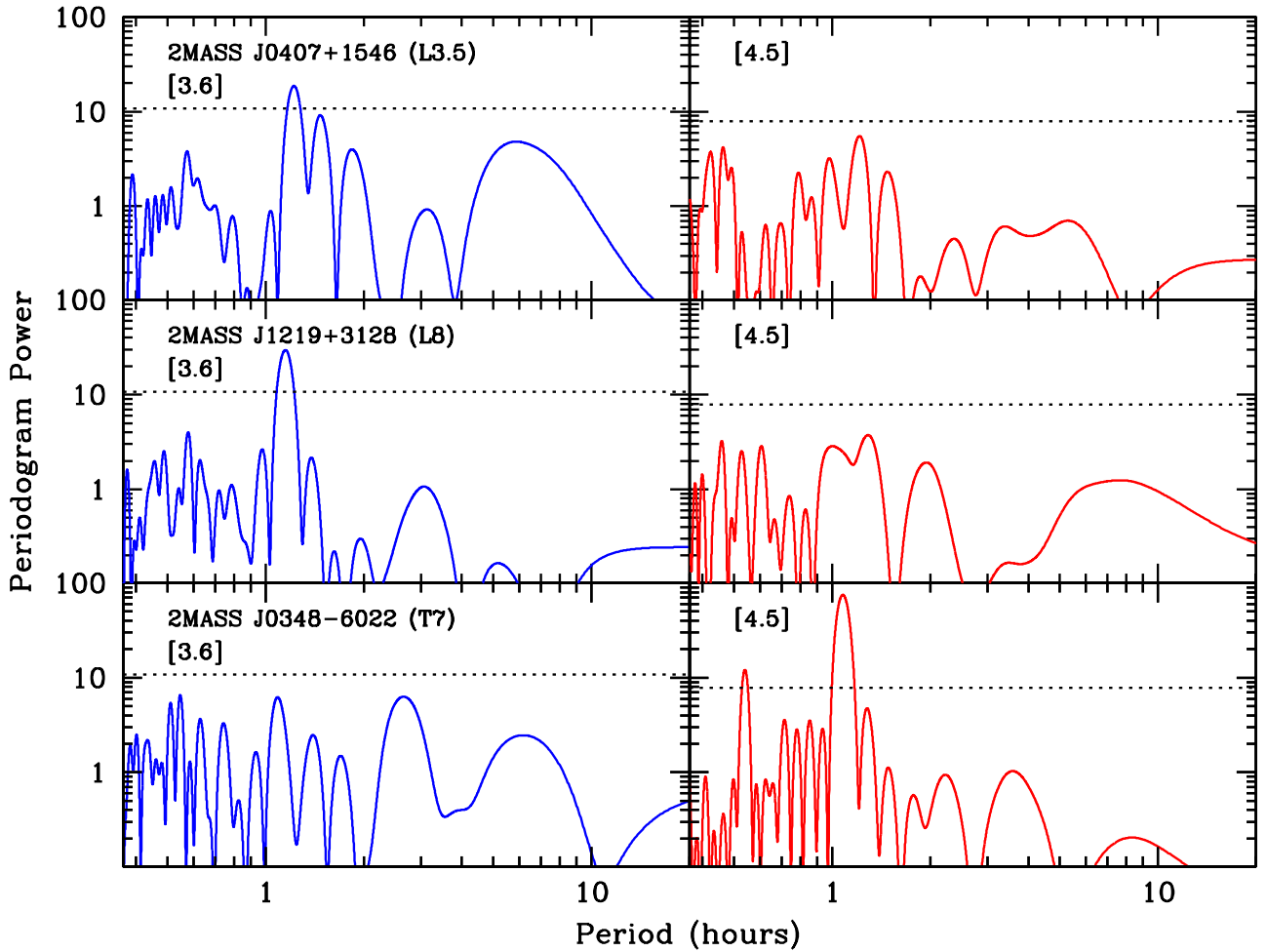}
  \caption{Lomb-Scargle periodogram power distributions of the light curves of our three targets for both Spitzer channels after the preliminary pixel phase correction (Sec.~\ref{sec:spitzer_variability}).
  We use the 95 percentile $p$-value thresholds determined from field stars (Fig.~\ref{fig:FAPvsmag}) to identify significant periodogram peaks. The relevant thresholds (dotted lines) at [3.6] and [4.5] are at periodogram powers of $P_{[3.6]}=10.7$ and $P_{[4.5]}=7.9$. 
  }
  \label{fig:periodogram}
\end{figure*}

The most significant periodogram peaks above the $p$-value threshold for our three targets are in the 1.1--1.2 hour range (Fig.~\ref{fig:periodogram}). In all three cases, significant periodicity is detected in only one of the two Spitzer IRAC channels: at [3.6] for the L dwarfs 2MASS~J0407+1546 and 2MASS~J1219+3128 and at [4.5] for the T dwarf 2MASS~J0348--6022.

At this stage of our analysis, the applied pixel phase correction is not in the final form presented in Section~\ref{sec:spitzer_periods}. The preliminary periodicities are potentially affected by Spitzer's known pointing `wobble.' The telescope's boresight follows a small sawtooth quasi-periodic oscillation with a mostly sub-hour time scale: the result of heater cycling to maintain adequate battery temperature \citep{Grillmair2012,Grillmair2014}. The amplitude of the pointing oscillation, up to 0.4~pix, can be sufficiently high to impact photometric measurements because of the pixel phase effect. During 2015, when our observations were taken, the mean pointing oscillation period was 49 minutes, with an inner quartile range of 43 minutes to 54 minutes \citep{Krick2018}. However, a small fraction of year 2015 observations in the Spitzer archive have pointing oscillation periods up to 80 minutes, similar to the 1.1--1.2 hour-long periods identified in our periodograms (Fig.~\ref{fig:periodogram}).

We do not believe that Spitzer's pointing wobble is responsible for the detected 1.1--1.2 hour periodicities for three reasons.  First, we expect roughly similar pixel phase-induced behavior of all point sources in our target fields.  Therefore, by setting a global 95\% $p$-value threshold in our preliminary analysis, we select for variability beyond what may be incurred by the pointing wobble.  Second, in Section~\ref{sec:spitzer_periods} we describe a more sophisticated photometric analysis that includes an astrophysical variability and a pointing oscillation model, and we clearly identify the wobble separately from the astrophysical periods. Finally, in Section~\ref{sec:SpectraAnalysis} we confirm that the rapid rotations implied by such short periods are expressed as wide Doppler line broadening in moderate-dispersion spectra of our three science targets.

\subsection{Simultaneous Fitting for Pixel Phase and Astrophysical Variability}
\label{sec:spitzer_periods}

\begin{figure*}
  \centering
    \includegraphics[trim={0.7cm 5.5cm 0.9cm 4.2cm},clip,width=0.95\textwidth]{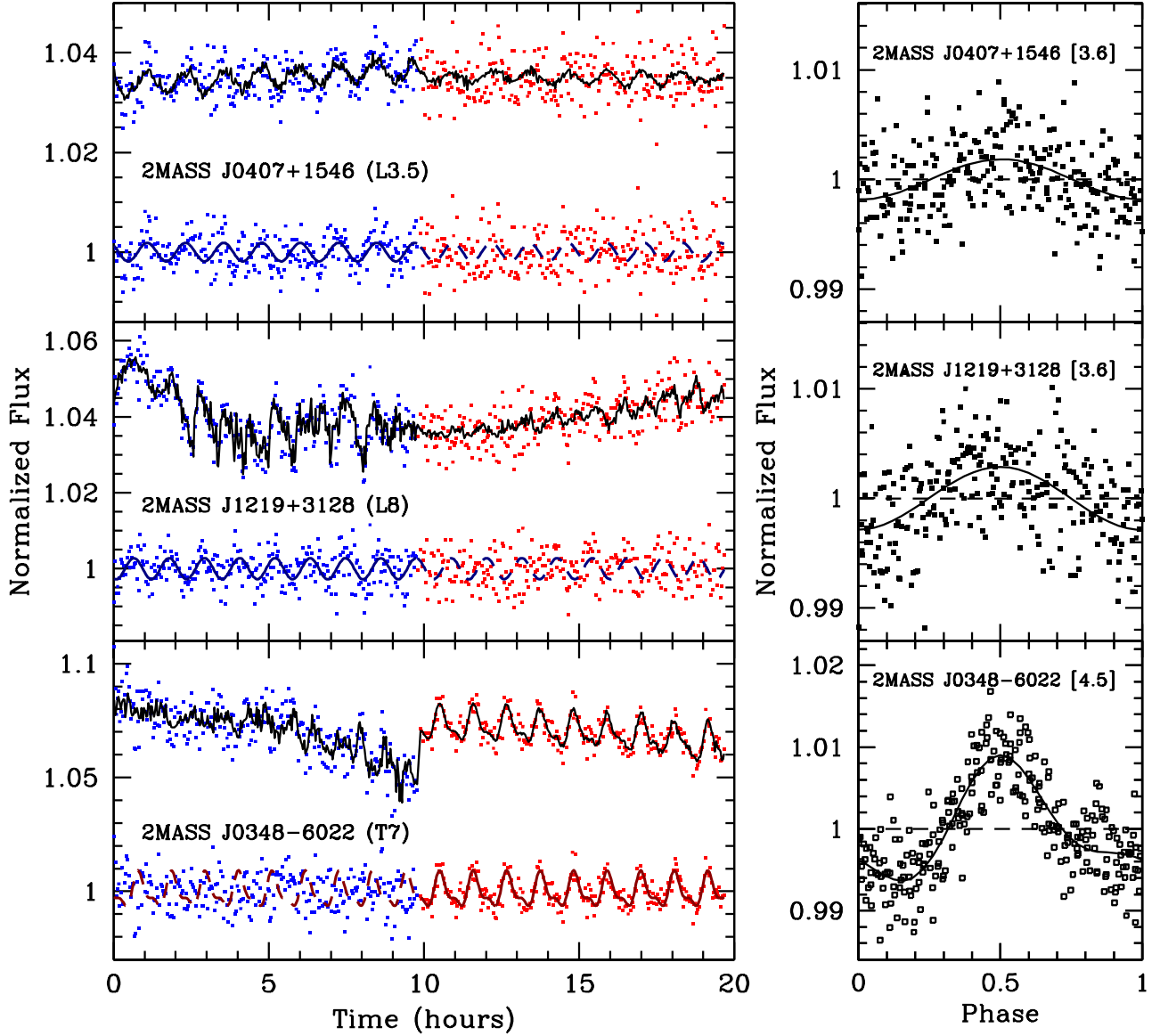}
  \caption{{\it Left:} Spitzer [3.6]- (blue) and [4.5]-band (red) light curves. Each target is shown in a separate panel, with the raw data on top. The bottom sequences show the final light curves after correcting for the pixel phase effect.  All light curves are normalized to unity, and the raw data are offset by a constant for clarity. A combined astrophysical and pixel phase model fit (Sec.~\ref{sec:spitzer_periods}) to the raw data is shown in black. The astrophysical model fits to the corrected data are shown in blue for [3.6] and red for [4.5]. The models are shown with a solid line over the channel that exhibits significant variability and with a dashed line over the other channel that exhibits no significant variability.
  {\it Right:}~Period-folded light curves in the channels with significant variability after the pixel phase correction. The mean flux level is represented as a horizontal dashed line at unity flux. The astrophysical model fit is shown as a solid line.
}
\label{fig:lightcurves}
\end{figure*}

Having identified candidate variables among the science targets with approximate photometry, we iterated our variability assessment with higher-precision photometry. We determined the optimal aperture for each object by seeking the lowest root-mean-square scatter in the measured fluxes. The optimal apertures in the [3.5]- and [4.5]-band data respectively were 1.4 and 2.1 pixels for 2MASS~J0348--6022, 1.4 and 2.0 pixels for 2MASS~J1219+3128, and 1.5 and 2.1 pixels for 2MASS~J0407+1546. We binned the photometry in groups of 10 consecutive measurements to lower the random noise. Our binning interval of 120 s still ensures fast enough sampling to retain sensitivity to the hour-long timescales of interest. We incorporated the initial period estimates from Section~\ref{sec:spitzer_variability} in an iterative least-squares method to simultaneously fit an astrophysical model (a truncated Fourier series) and a correction for the pixel phase effect in both channels \citep{Heinze2013}.  We show the raw and corrected Spitzer light curves for our three targets in the left panel of Figure~\ref{fig:lightcurves}.

The raw photometry in Figure~\ref{fig:lightcurves} shows the sawtooth pointing oscillation of Spitzer in the light curves of two of the three science targets. The effect is present mostly throughout the [3.6]- and [4.5]-band staring observation of 2MASS~J1219+3128 (middle-left panel of Fig.~\ref{fig:lightcurves}), with a sawtooth-like pattern that repeats 10 times over 10 hours at [3.6]. The corresponding 60 minute time scale of the sawtooth pattern is distinct from the 68 minute astrophysical period seen in the corrected [3.6]-band light curve. The latter half of the [3.6]-band observation of 2MASS~J0348--6022 also shows sawtooth variations on a 60 minute time scale. However, no astrophysical variability is detected in 2MASS~J0348--6022 at [3.6].  This T7 dwarf is variable only at [4.5], where no effect of the sawtooth pattern is seen.

\begin{figure}
  \centering
    \includegraphics[trim={0.7cm 5.5cm 4.5cm 4.cm},clip,width=0.75\textwidth]{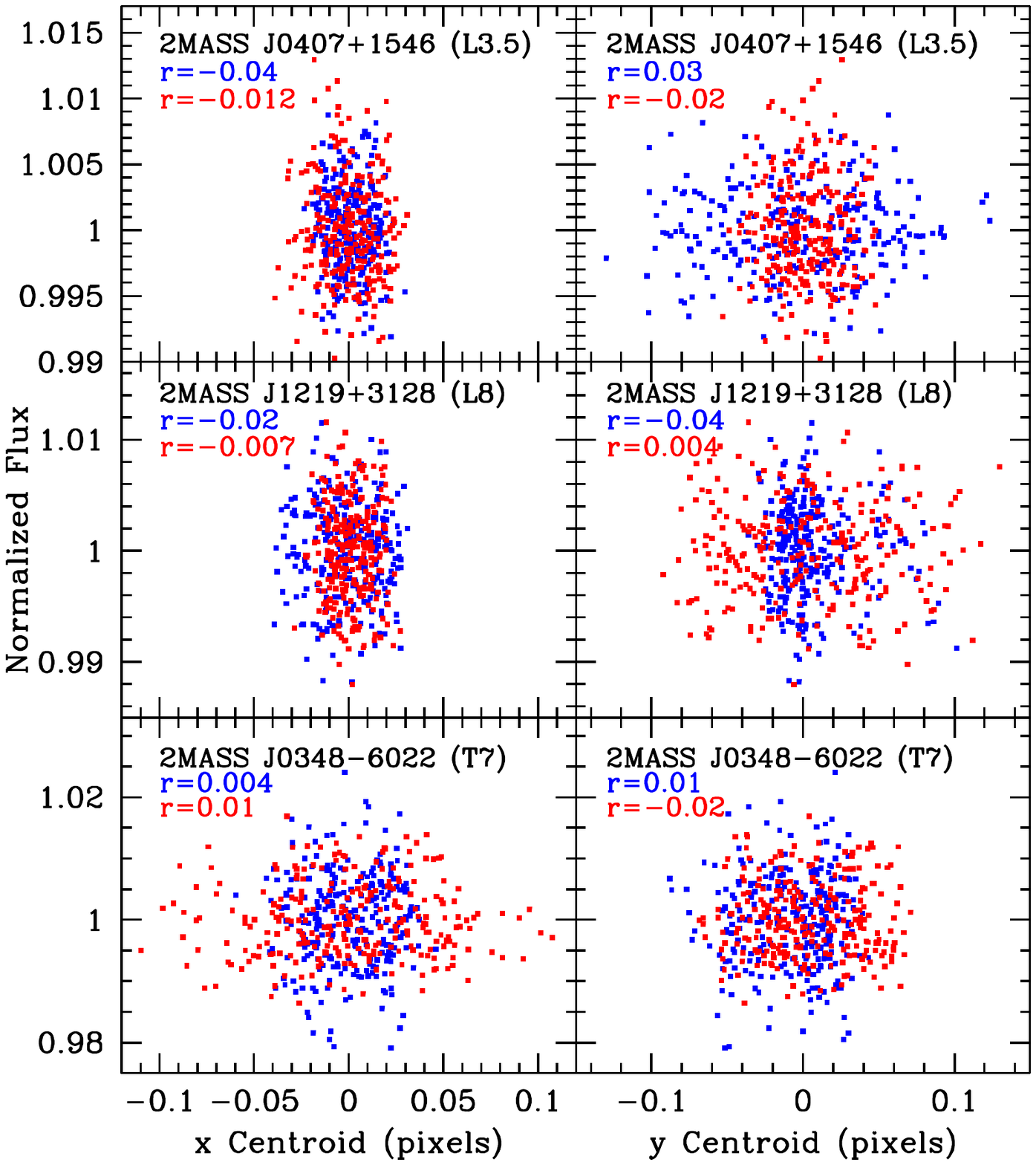}
  \caption{Pixel phase-corrected flux at [3.6] (blue) and [4.5] (red) as a function of centroid position in both the $x$- and $y$-directions. The centroids are measured relative to the average centroids across all exposures. The Pearson correlation coefficients ($r$) are given in each panel and we find that there is no correlation between the flux and centroid positions on the detector. We conclude that there is no residual periodic effect on the photometry after correcting for Spitzer's pointing wobble.
}
\label{fig:centroids}
\end{figure}

We further verified that there is no residual periodicity effect from the pointing wobble by confirming that there is no correlation between the flux and centroid position on the detector after correcting our photometry for pixel phase (Fig.~\ref{fig:centroids}). We computed Pearson correlation coefficients of $|r| \leq 0.04$ between flux and centroid position for each object and Spitzer channel. We conclude that Spitzer's pointing oscillation is not the cause of the variability we observe.

We adopt the results from the simultaneous astrophysical and pixel phase model as the true periods and peak-to-trough amplitudes of our variables, rather than the preliminary results from the periodogram fitting shown in Figure~\ref{fig:periodogram}. In all three cases the final and the preliminary periods agree to within 1\%, and the periodogram power of the significant peaks increased for the final, corrected data. From our best-fit astrophysical model, the two L dwarfs require only a single Fourier term for an adequate light-curve fit. The T7 dwarf 2MASS~J0348--6022 requires a two-term Fourier fit, and so both significant peaks seen in the [4.5]-periodogram (Fig.~\ref{fig:periodogram}) are astrophysical in nature.  However, the higher-frequency oscillation is less significant, and is a harmonic at half the period: 0.54 h versus 1.08 h. It may indicate a two-spot configuration on opposite hemispheres of the T dwarf.

We use a Markov chain Monte Carlo (MCMC) analysis, as described in Section 3.4 of \citet{Heinze2013}, to determine the uncertainties on the periods and the amplitudes.  We fit the [3.6] and [4.5] photometry simultaneously, by requiring the same period but different amplitudes for the two channels. Since in all cases only one of the channels shows significant variability, we set 2$\sigma$ upper limits on the amplitude ratios of the ``non-variable'' channels to the variable channels.

The periods of the T7, L3.5, and L8 dwarfs range between 1.08 h and 1.23 h: faster than any measured before (see Section~\ref{sec:Discussion}).  We show the phase-folded light curves in the variable channel for each target in the right panel of Figure~\ref{fig:lightcurves}.  The object names, spectral types, magnitudes, variable channels, and photometric periods, peak-to-trough amplitudes, and amplitude ratios are listed in Table~\ref{table:fastperiods}. \\

\begin{deluxetable}{l c c c c c c c c} 
\tabletypesize{\scriptsize}
\tablecolumns{10} 
\tablewidth{0pt}
\tablecaption{Spitzer Photometry and Results from Markov Chain Monte Carlo Analysis of Periods and Peak-to-trough Amplitudes \label{table:fastperiods}}
\tablehead{
\colhead{Object}\vspace{-0.2cm} & \colhead{Spectral} & \colhead{Date} & \colhead{[3.6]} & \colhead{[4.5]} & \colhead{Variable}& \colhead{Period\tablenotemark{c}}& \colhead{Amplitude} & \colhead{Amplitude} \\ 
\colhead{}\vspace{-0.2cm} & \colhead{Type\tablenotemark{a}} & \colhead{Observed } & \colhead{(mag)} & \colhead{(mag)}  & \colhead{Channel\tablenotemark{b}} & \colhead{(h)} & \colhead{in Variable} & \colhead{Ratio\tablenotemark{d}} \\
\colhead{} &  \colhead{} & \colhead{} & \colhead{}  & \colhead{} & \colhead{} & \colhead{}  &  \colhead{Channel\tablenotemark{c} (\%)} & \colhead{}
}
\startdata 
2MASS J04070752$+$1546457 & L3.5 & 2015 Apr 26 & $12.83 \pm 0.01$ & $12.91 \pm 0.01$ & [3.6] & $1.23~[1.22, 1.24]$ & $0.36~[0.24, 0.46]$ & $<$1.2\\
2MASS J12195156$+$3128497 & L8 & 2015 Sep 16 & $13.40 \pm 0.02$ & $13.26 \pm 0.02$ & [3.6] & $1.14~[1.13, 1.17]$ & $0.55~[0.42, 0.69]$ & $<$0.53\\
2MASS J03480772$-$6022270 & T7 & 2015 Apr 22  & $14.36 \pm 0.03$ & $12.86 \pm 0.02$ & [4.5] & $1.080~[1.075, 1.084]$ & $1.5~[1.4, 1.7]$ & $<$0.58\\
\enddata
\tablenotetext{a}{Spectral type references, in row order: \citet{Reid2008}, \citet{Chiu2006}, \citet{Burgasser2003}.}
\tablenotetext{b}{Each target varies in only one of the two Spitzer IRAC channels.}
\tablenotetext{c}{Square brackets denote the 2$\sigma$ confidence intervals on the periods and aplitudes determined from our MCMC analysis (Sec.~\ref{sec:spitzer_periods}).}
\tablenotetext{d}{This is the 2$\sigma$ upper limit on the amplitude ratio between the two channels. The ratio is of the non-variable channel to the variable channel.}
\end{deluxetable}

\subsection{Discussion of Photometric Variability: Periods and Mechanisms}
\label{sec:SpitzerDiscussion}

Two of our three targets have been previously reported as potential variables. For 2MASS~J1219+3128 (L8), \citet{Buenzli2014} find a lower limit of 2\% on the variability amplitude in a 1.12--1.20 \micron\ subset of their 1.1--1.7 \micron\ HST/WFC3 spectra, over a 36 minute sequence of nine spectroscopic exposures. However, the variability is not significant over any other part of their 1.1--1.7 \micron\ spectra, and they classify the detection as tentative.  For 2MASS~J0348$-$6022 (T7), \citet{Wilson2014} report a $J$-band amplitude of $2.4\%\pm0.5\%$ in a three hour long observation.  However, a re-analysis of their NTT/SofI observations by \citet{Radigan2014b} shows that the reported variability is likely spurious, and related to residual detector and sky-background systematics. \citet{Radigan2014b} revised the $J$-band variability in the \citet{Wilson2014} observations to a $< 1.1\%\pm0.4\%$ upper limit. Similarly, a 1\% upper limit for 2MASS J0348$-$6022 is deduced from a prior six hour $J$-band monitoring observation by \citet{Clarke2008}, also with NTT/SofI.  No variability has been previously reported for 2MASS~J0407+1546 (L3.5). 

The small periodogram $p$-values and large periodogram powers (Figs.~\ref{fig:FAPvsmag} and \ref{fig:periodogram}) of our Spitzer observations confidently establish that all three L and T dwarfs exhibit periodic variability. Each of our three targets varies in only one of the two IRAC channels within the photometric precision limits. The two L dwarfs vary only at [3.6], whereas the T7 dwarf varies only at [4.5]. Such behavior is consistent with prior observations of infrared variability trends with spectral type. \citet{Metchev2015} found that five of their 19 variable L3--T8 dwarfs varied only at [3.6] (two L3s, two L4s, and a T2), and one (T7) dwarf varied only at [4.5].  Single-band [3.6] variations in an L dwarf have also been reported by \citet{Gizis2015}, while [4.5]-only variations are seen in Y dwarfs \citep{Cushing2016, leggett2016}. 

Wavelength-dependent amplitude differences are explained by the dominant gas absorption species in the atmosphere. In wavelength regions of strong molecular gas opacity, clouds reside below the photosphere and so cloud heterogeneities are obscured. Cloud structures are detectable only in relatively transparent spectral regions, away from dominant molecular bands \citep[e.g.,][]{Ackerman2001}. With CO being a dominant source of upper-atmosphere gas opacity in L dwarfs, cloud condensate-induced variability will be suppressed around the 4.5~$\mu$m fundamental CO band (i.e., in IRAC channel 2). Conversely, variability around the 3.3~$\mu$m CH$_4$ fundamental band (within IRAC channel 1) will be suppressed in T dwarfs.

Alternative variability mechanisms that do not require clouds have also been proposed.  Such scenarios do not imply that clouds may not exist at all in the atmospheres of brown dwarfs, just that they are not responsible, or not entirely responsible, for the observed variability.  For example, some variable brown dwarfs show radio emission that may be best explained as auroral in nature (e.g., \citealt{Antonova2008,Hallinan2015,Kao2018}).   \citet{Richey-Yowell2020} correlate such auroral signatures with the presence of H$\alpha$ emission.  One of our three variables, the L3.5 dwarf 2MASS~J0407+1546, is a strong H$\alpha$ emitter \citep[equivalent width of 60\AA;][]{Reid2008}.  While \citet{Miles2017b} find no correlation between H$\alpha$ emission and large-amplitude ($\gtrsim 1\%$) variations (a result also confirmed by \citealt{Richey-Yowell2020}), the more subdued 0.36\% variation in 2MASS~J0407+1546 could well be magnetic in origin.

\citet{Robinson2014} propose atmospheric temperature fluctuations as a potential cause of photometric variability. They show that thermal perturbations occuring deep in the atmosphere can cause surface brightness fluctuations at infrared wavelengths. \citet{Tremblin2015,Tremblin2020} show that fingering convection in a cloudless atmosphere can also result in variability. 
Ultimately, the observations that we present are not decisive of the variability mechanism, and our focus is instead on the short rotation periods.

The periodic regularity seen in the light curves of our three targets (Fig.~\ref{fig:lightcurves}) argues for one (2MASS~J0407+1546, 2MASS~J1219+3128) or two (2MASS~J0348$-$6022) dominant photospheric spots. An alternative interpretation of these data is that we are seeing a repeating spot pattern extended along a band on a more slowly rotating object, e.g., as in the case for Jupiter \citep{dePater2016}. Additionally, \citet{Apai2017} show that the variability of infrared brightness in T dwarfs can be dominated by planetary-scale waves. They find that the combined variability effect of multiple sets of planetary waves or spots may place the periodogram peak at half the true period, or that double peaks may occur near the true rotation period due to differential rotation.

To the sensitivity of our data, none of our objects show the kind of complex light modulations seen in the \citet{Apai2017} T dwarfs.  However, Jupiter-like repeated spot patterns remain a possibility.  In Sections~\ref{sec:spectra} and \ref{sec:SpectraAnalysis} we use near-infrared spectroscopy to measure the projected rotation velocities $v\sin i$ of our targets and confirm that all three rotate rapidly.

\section{Spectroscopic Observations}
\label{sec:spectra}

If our objects are truly rapidly rotating, then their spectroscopic line profiles will be significantly Doppler-broadened, while more slowly rotating objects with repeating spot patterns will not show much line broadening.  Herein we report $R=6000 - 12,000$ near-infrared spectroscopy which we use to confirm the rapid rotations and in Section~\ref{sec:SpectraAnalysis} to estimate the objects' fundamental parameters.

We present a previously unpublished spectrum of the T7 dwarf 2MASS J0348$-$6022 and a new observation of the L8 dwarf 2MASS J1219$+$3128 at a resolution of 6000 over 0.91--2.41 $\mu$m with the Folded-port InfraRed Echellette (FIRE; \citealp{Simcoe_etal2008, Simcoe_etal2013}) at the Magellan Baade telescope. We also observed the L3.5 dwarf 2MASS J0407$+$1546 at a resolution of 12,000 over 2.275--2.332 $\mu$m with the Gemini Near-InfraRed Spectrograph (GNIRS; \citealt{Elias2006}) at the Gemini North Observatory. The spectroscopic observations are summarized in Table~\ref{table:observations}.

\begin{deluxetable}{l c c c c c c c c c} 
\tablecolumns{10} 
\tablewidth{0pt}
\tablecaption{Magellan/FIRE and Gemini North/GNIRS Spectroscopic Observations. \label{table:observations}}
\tablehead{
\colhead{Target}\vspace{-0.2cm} & \colhead{$K_s$} & \colhead{Date} & \colhead{Instrument} & \colhead{Resolution} & \colhead{Exposure}& \colhead{S/N}& \colhead{Target} & \colhead{Telluric}&  \colhead{Telluric} \\
\colhead{}\vspace{-0.2cm} & \colhead{(mag)} & \colhead{Observed} & \colhead{} & \colhead{}  & \colhead{Time} & \colhead{} & \colhead{Airmass} & \colhead{Standard} & \colhead{Standard}\\
\colhead{} & \colhead{} & \colhead{} & \colhead{} & \colhead{}  & \colhead{(minutes)} & \colhead{} & \colhead{} & \colhead{} & \colhead{Airmass}
}
\startdata 
2MASS J03480772$-$6022270 & 15.60 & 2012 Jan 3 & Magellan/FIRE & 6000 & 30.3 & 36 & 1.25--1.27 & HD 28667 & 1.28\\
2MASS J12195156$+$3128497 & 14.31 & 2017 Feb 16 & Magellan/FIRE & 6000 & 26.6 & 46 & 2.12--2.31 & HD 96781 & 1.98\\
2MASS J04070752$+$1546457 & 13.56 & 2017 Oct 10 & Gemini North/GNIRS & 12,000 & 80.0 & 29 & 1.07--1.28 & HD 17971 & 1.03\\
\enddata
\tablecomments{$K_s$ magnitudes are from 2MASS \citep{Cutri2003}. The signal-to-noise ratio is the median around the $K$-band peaks of the FIRE spectra of for 2MASS~J0348$-$6022 (between 2.05 and 2.15 $\mu$m) and 2MASS~J1219$+$3128 (between 2.1 and 2.2~$\mu$m), and the median over the full range of the GNIRS spectrum of 2MASS~J0407$+$1546.
}
\end{deluxetable}

\subsection{Magellan/FIRE Spectroscopy: 2MASS J0348$-$6022 (T7) and 2MASS J1219$+$3128 (L8)}

For our FIRE observations we used the cross-dispersed echelle mode with the 0$\farcs$6 (3.3 pixel) slit aligned to the parallactic angle to obtain $R\approx6000$ spectra over 0.91--2.41 $\mu$m. We observed 2MASS J0348$-$6022 on 2012 January 3 (UT) under clear skies with 0$\farcs$7 $J$-band seeing and airmass of 1.25--1.27. We obtained two 909~s exposures, dithered along the slit. We observed the A0~V star HD~28667 ($V = 6.87$ mag) in four 1~s dithered exposures following the 2MASS J0348$-$6022 observations at a similar airmass (1.28). We observed 2MASS J1219$+$3128 on 2017 February 16 (UT) under clear skies with 1$\farcs$2--1$\farcs$4 $J$-band seeing and airmass of 2.12--2.31. We obtained four 400~s exposures dithered pair-wise along the slit. We observed the A0~V star HD~96781 ($V = 10.2$ mag) in six 1~s dithered exposures at a similar airmass (1.96--2.05). For both sets of observations we obtained ThAr emission lamp spectra after each target. We obtained dome and sky flat-field observations at the beginning of each night for pixel response and slit illumination calibration.

\begin{figure}
  \centering
    \includegraphics[trim={1.65cm 5cm 0.25cm 2.5cm},clip,width=0.95\textwidth]{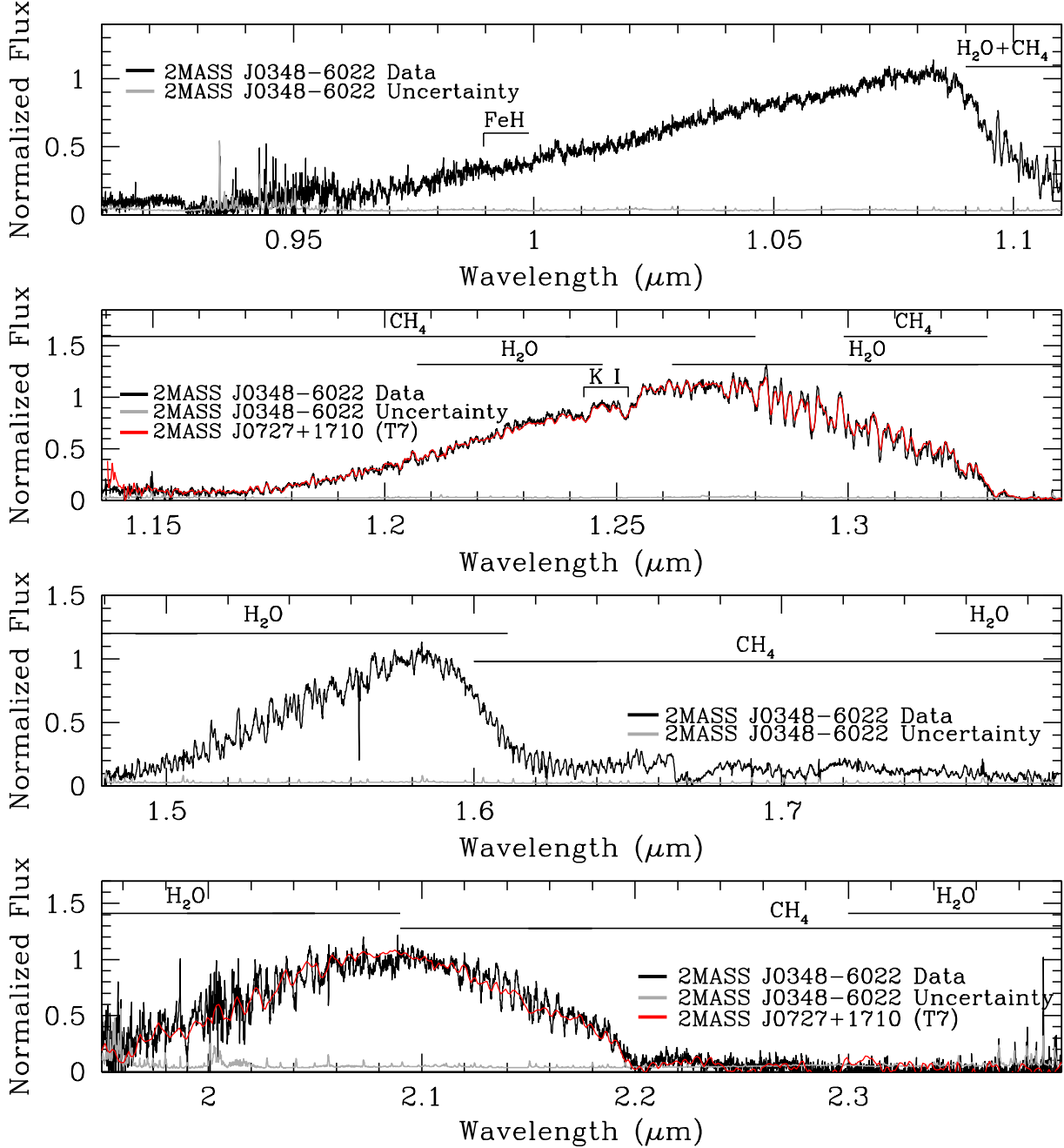}
  \caption{Magellan/Folded-port InfraRed Echellette (FIRE) $z$, $J$, $H$, and $K$-band spectra (from top to bottom) of 2MASS J0348$-$6022 (black), compared to template T7 spectra where available. The uncertainty on the FIRE spectrum is shown in gray along the bottom of each panel. The template spectrum of 2MASS J07271824+1710012 is from the Brown Dwarf Spectroscopic Survey \citep{BDSS2003}. Major molecular features \citep{BDSS2003, cushing_etal06, Bochanski2011} are indicated.
  }
  \label{fig:0348templates}
\end{figure}

\begin{figure}
  \centering
    \includegraphics[trim={1.65cm 5cm 0.25cm 2.5cm}, clip,width=0.95\textwidth]{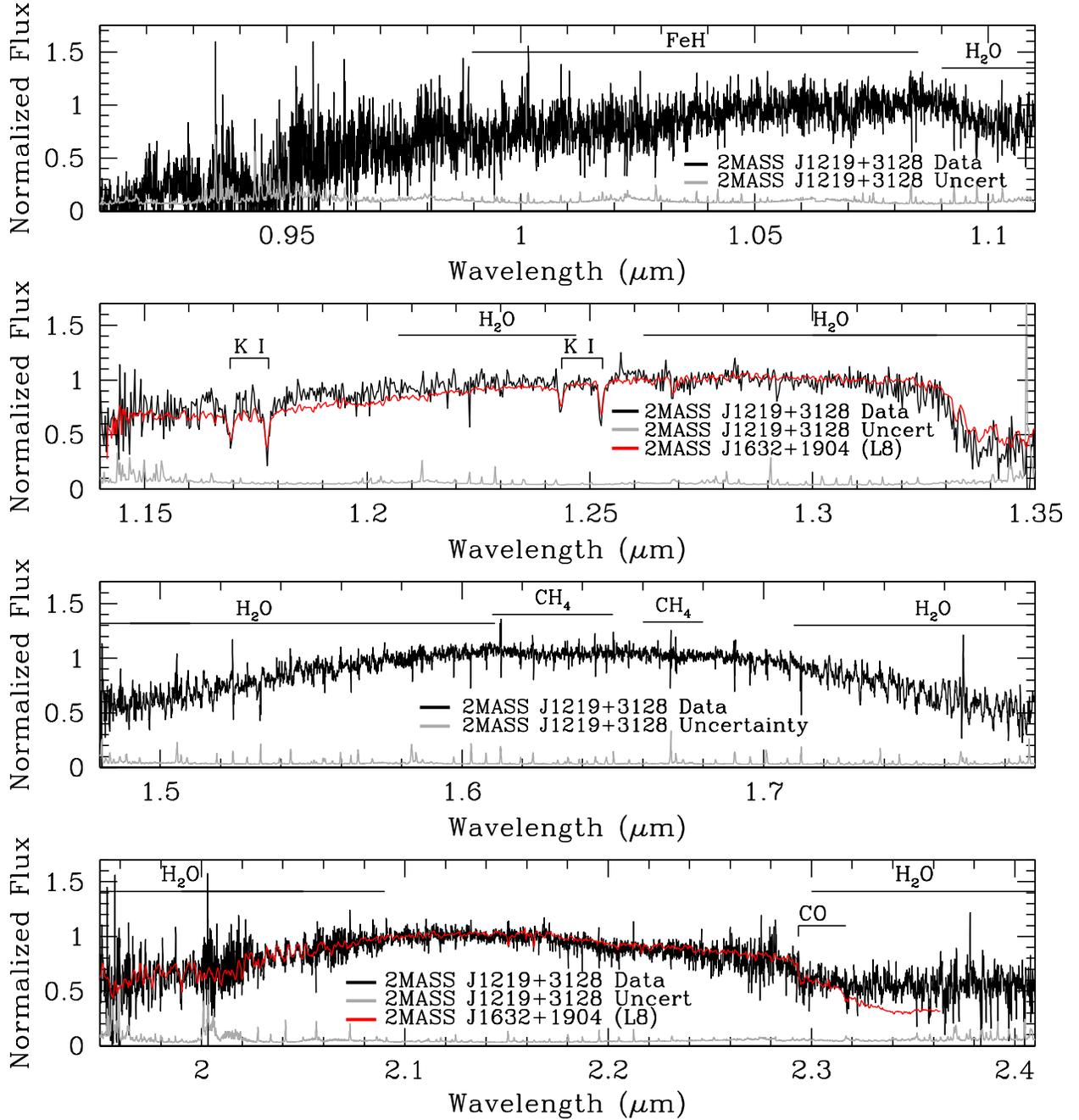}
  \caption{Magellan/FIRE $z$-, $J$-, $H$-, and $K$-band spectra (from top to bottom) of 2MASS J1219$+$3128 (black), compared to template L8 spectra where available. The uncertainty on the FIRE spectrum is shown in gray along the bottom of each panel. The template spectrum of 2MASS J16322911$+$1904407 is from the Brown Dwarf Spectroscopic Survey \citep{BDSS2003}. Major molecular features \citep{BDSS2003, cushing_etal06} are indicated.
   }
  \label{fig:1219templates}
\end{figure}

The FIRE data were reduced using the Interactive Data Language ({\sc IDL}) pipeline FIREHOSE v2 \citep{Gagnezenodo},  which is based on the MASE \citep{Bochanski_etal2009} and SpeXTool \citep{Vacca_etal2003, Cushing_etal2004} packages.\footnote[15]{\url{https://github.com/jgagneastro/FireHose_v2/}} 
Details for standard reduction of point-source data with FIREHOSE are described in \citet{Bochanski2011}. The ThAr lamp images were used to trace the spectral orders and derive pixel response and illumination corrections which were applied to the science frames. A combination of OH telluric lines in the science frames and ThAr emission lamp lines were used to determine the wavelength solution along the center of each order and the order tilt along the spatial direction, so as to construct a two-dimensional vacuum wavelength map. The typical uncertainty of the wavelength solution was 0.20 pixels, corresponding to a precision of 3.0 km\,s$^{-1}$. The sky background in each frame was fit with a two-dimensional sky model constructed using basis splines \citep{Kelson2003}, which was then subtracted from the frame. One-dimensional spectra were optimally extracted \citep{Horne1986} in each order onto a heliocentric wavelength frame. Correction for telluric absorption and overall flux calibration was determined from the A0~V star spectra using a modified version of \texttt{xtellcor} from SpeXtool \citep{Vacca_etal2003, Cushing_etal2004}. Spectra from the individual frames of the FIRE data were combined for each target after relative flux normalization, and the individual orders were merged into one-dimensional spectra. The resulting reduced spectra are shown in Figure~\ref{fig:0348templates} for 2MASS J0348$-$6022 and in Figure~\ref{fig:1219templates} for 2MASS J1219$+$3128, along with comparison spectra, where data at similar resolution of other objects of the same spectral types were available from the literature.

\subsection{Gemini/GNIRS Spectroscopy: 2MASS J0407$+$1546 (L3)}

\begin{figure}
  \centering
    \includegraphics[trim={1.65cm 16cm 0.25cm 3.25cm},clip,width=0.9\textwidth]{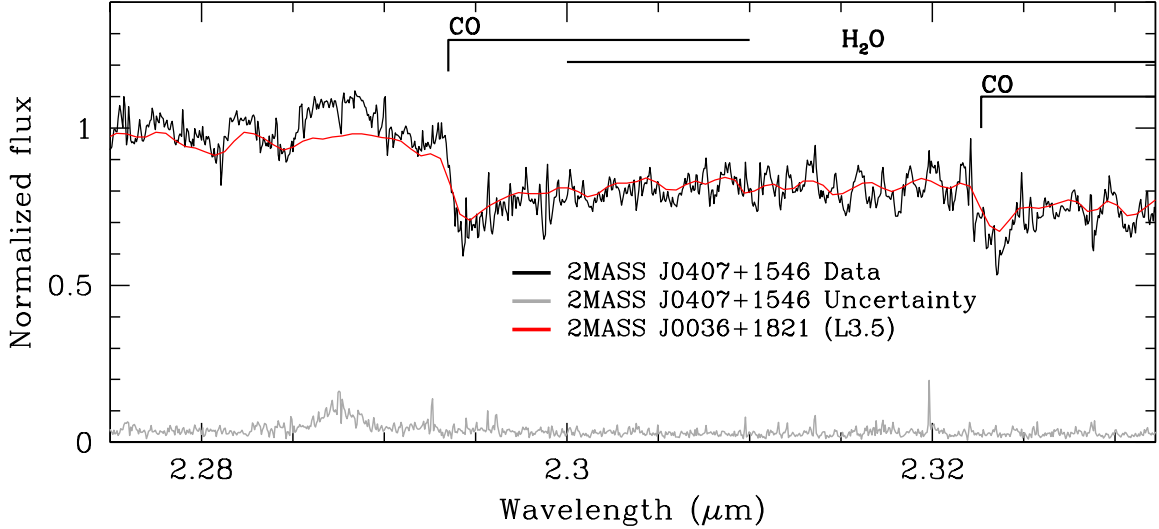}
  \caption{Gemini/Gemini Near-InfraRed Spectrograph (GNIRS) $K$-band spectrum of 2MASS J0407$+$1546 (black), compared to a template L3.5 spectrum \citep[2MASS J00361617$+$1821104 from IRTF/SpeX;][]{Rayner2009}. The uncertainty on the GNIRS spectrum is shown in gray along the bottom of the panel. Major molecular features \citep{cushing_etal06} are indicated.
  }
  \label{fig:0407templates}
\end{figure}

Our GNIRS observations of the L3.5 dwarf 2MASS J0407$+$1546 took place on 2017 October 10 (UT). We followed the same procedure and instrument settings as used by \citet{allers_etal16} for radial and rotation velocity measurements of an L dwarf. We used the 111 lines mm$^{-1}$ grating with the 0$\farcs$15 (3.0~pixels) slit aligned to the parallactic angle to obtain $R\approx 12,000$ 2.27--2.33 $\mu$m spectra at an airmass of 1.07--1.28. We obtained eight 600~s exposures, dithered between two positions on the slit. We observed the A0~V star HD~17971 ($V = 8.78$ mag) for telluric absorption correction in eight 60~s dithered exposures at a similar airmass (1.03) using the same instrument setup. ThAr emission lamp observations were obtained immediately after the 2MASS~J0407$+$1546 observations. The 2MASS~J0407$+$1546 data were reduced using a combination of general and Gemini-specific IRAF\footnote[16]{Image Reduction and Analysis Facility, distributed by the National Optical Astronomy Observatories.} routines. Data were prepared, sky-subtracted, and flat-fielded using the Gemini tasks \texttt{nsprepare}, \texttt{nsreduce}, and \texttt{nsflat}. Individual spectra were extracted using \texttt{apall}. The XeAr lamp spectrum was extracted once for each of the science spectra, using the science extraction traces as references, and \texttt{identify} and \texttt{dispcor} were used to identify calibration lines and generate a wavelength solution for each science spectrum. A Legendre polynomial was used with \texttt{identify}, typically of second order. The typical uncertainty in the wavelength calibration was 0.25~pixels, which corresponds to a precision of 2.0~km~s$^{-1}$. The individual wavelength-calibrated spectra were median-combined, and the standard deviation was adopted as the uncertainty. The same reduction steps were repeated for the standard star, and the science spectrum was divided by the standard spectrum to remove telluric lines, and multiplied by a $T_{\rm eff}=9600$ K blackbody. The resulting spectrum is shown in Figure~\ref{fig:0407templates}, along with a comparison spectrum of another L3.5 dwarf.

\section{Confirmation of Rapid Rotations and Determination of Physical Parameters}
\label{sec:SpectraAnalysis}

We compared our spectra to the photosphere models of \citet[SM08]{Saumon2008}, \citet[BT-Settl]{Allard2012}, \citet[Morley]{Morley2012}, and \citet[Sonora]{Marley_Sonora_spectra} to determine the physical properties of our objects. All but the BT-Settl models are based on the \citet{Ackerman2001} cloud model. The model photospheres are provided on fixed grids of effective temperature ($T_{\rm eff}$) and surface gravity ($\log g$). The SM08 and Morley models also have a sedimentation efficiency ($f_{\rm sed}$) parameter. The $T_{\rm eff}$ grids are in steps of 100~K for all except the BT-Settl models, which are in 50~K steps, and the $\log g$ grids are in steps of 0.5~dex for all except the Sonora models, which are in steps of 0.25~dex.

We also implemented grids for radial velocity (RV) and $v \sin i$ in steps of 0.1 km\,s$^{-1}$. For the RV we applied a simple Doppler shift to the wavelength of the models. For $v \sin i$ we simulated rotational broadening by convolving the model spectra with the standard rotation kernel from \citet{Gray1992} using the \texttt{lsf\_rotate} task in the {\sc IDL} Astronomy User's Library.\footnote[17]{\url{https://idlastro.gsfc.nasa.gov/ftp/pro/astro/lsf\_rotate.pro}} 

We first verified the spectral types of our targets by overlaying the spectra of other well-studied L and T dwarfs (Figs.~\ref{fig:0348templates}--\ref{fig:0407templates}). For each object we restricted the effective temperature grids to 300~K above and below the expected values for each spectral type based on \citet{Filippazzo2015}. We did not restrict the $\log g$ and $f_{\rm sed}$ (where available) grids.

The quality of the model fits to the full-band FIRE spectra is dominated by the low-order continuum which is mostly affected by the effective temperature and, when using the SM08 and Morley models, the sedimentation efficiency. Instrument systematics may also affect the continuum shape of our Magellan/FIRE spectra that cover a wide (0.91--2.41~$\mu$m) wavelength range. Broadband model fits thus preclude us from obtaining accurate information about the RV and $v\sin i$, both of which do not depend on the continuum but entirely on the positioning and profiles of spectral lines.  The pressure-broadening effect of surface gravity is also well reflected in the theoretical line profiles, even though surface gravity does affect the continuum of model ultra-cool photospheres. 

To extract accurate estimates of RV, $v\sin i$, and $\log g$, we fit models to select narrow-wavelength sub-regions of the FIRE spectra that are dominated by dense sequences of H$_2$O, FeH, or CH$_4$ absorption lines, as marked in Figures \ref{fig:0348templates} to \ref{fig:0407templates}.  It is likely that in doing so we may still be affected by wavelength systematics among the theoretical line lists for the different molecules.  In addition, the different wavelength sub-regions probe different atmospheric depths and pressures.  Hence, a wavelength region where flux originates deeper in the atmosphere could exhibit greater pressure broadening compared to a region where the flux originates higher up.  We account for these effects by selecting several different narrow-wavelength regions from the Magellan/FIRE spectra (see Table~\ref{table:modelfits}) and, as much as possible, different molecular absorbers. Overall, we find that the values for RV, $v\sin i$, and $\log g$ obtained from the narrow-wavelength regions are more self-consistent, with uncertainties 1.5--3 times smaller, than those from the full bands.

Our approach was first to fit each of the narrow regions to determine RV, $v \sin i$, and $\log g$ and then to fit the full bands ($z$: $0.91-1.11~\mu$m, $J$: $1.14-1.345~\mu$m, $H$: $1.48-1.79~\mu$m, and $K$: $1.96-2.35~\mu$m) to determine $T_{\rm eff}$ and $f_{\rm sed}$. 
We adopt the RV, $v\sin i$, and $\log g$ values determined from the narrow regions of the FIRE spectra of 2MASS~J0348--6022 (T7) and 2MASS~J1219+3128 (L8) as the fiducial values for these objects (Table~\ref{table:modelfits}).  While the narrow-band fits also produce estimates for $T_{\rm eff}$, the full-band spectra are likely more sensitive to it. Then, in re-applying the models to the full-band spectra, we allow RV and $v \sin i$ to vary only within 2$\sigma$ of the  values adopted from the narrow regions.  That is, we constrain RV and $v \sin i$ within a small range, as they should not effect the determinations of $T_{\rm eff}$ and (where applicable) $f_{\rm sed}$.  We still allow $\log g$ to be a free parameter in the full-band fitting because of its stronger effect on the continuum. This mirrors our approach for fitting the narrow regions, where we allow $T_{\rm eff}$ to be a free parameter, even if we adopt the results from the broad regions.  In this manner we probe the full parameter space for both $\log g$ and $T_{\rm eff}$ in each case, and obtain a more reliable estimate for each. Ultimately, the two sets of determinations for $T_{\rm eff}$ are consistent with each other (Tables~\ref{sec:spectra}--\ref{table:inclinations}). Estimates for $\log g$ tend to be 0.5--1.0~dex higher based on the line profile fits compared to the continuum fits in all models.  We favor the former, as they are closer to the fundamental radiative transfer calculations for each species.  The latter involve additional considerations of convection and relative chemical abundances.

The spectral range of our Gemini/GNIRS observation of 2MASS~J0407+1546 is much narrower, so we consider it only in its entirety. 

In terms of specific steps to fit models to the data, we started by normalizing the data to unity. In the narrow regions we divided by the median flux value, and for the full-band data, we divided by a constant such that the peak flux was unity. We shifted the model for radial velocity, broadened for $v \sin i$, and then smoothed the model to the resolution of the data. We also determined a flux zero-point to be added to the data and a multiplicative factor to scale the model which minimized the $\chi^2$ statistic ($\chi^2 = \sum_{i=1}^{N} [(O_i - M_i)^2 / \sigma_i^2]$, where $O_i$ is the observed flux, $M_i$ is the flux of the model, and $\sigma_i$ is the uncertainty of the data). We computed the offset, multiplicative factor, and $\chi^2$ statistic for every model on the grid of $T_{\rm eff}$, $\log g$, RV, $v \sin i$, and $f_{\rm sed}$ (where available).

The probability of a given $\chi^2$ value is 
$p \propto e^{-\chi^2/2}$. We computed the probabilities for every model on our grid, normalized the sum of the $p$-values to unity, then marginalized over each of the parameters to obtain the probability distributions for each parameter. The distributions for RV and $v \sin i$ were Gaussian in shape, and we report the mean values and 1$\sigma$ error bars for in Table~\ref{table:modelfits}. For the other parameters the model grid spacing was coarse, and the probability of values other than the values presented in Tables~\ref{table:modelfits} and~\ref{table:modelfitsbroad} is negligible. We report the results for these parameters with error bars corresponding to the grid-spacing. Tables~\ref{table:modelfits} and~\ref{table:modelfitsbroad} give the most probable values for each family of models in each wavelength region.

We find that the best-fit parameter values can vary significantly between model families and between different wavelength regions, while giving comparable reduced $\chi^2$ statistics. Understanding the subtle differences between the families of models is related to the molecular line lists and opacities used to compute these models, and is beyond the scope of this paper. 

To determine the final values of the parameters, we take a weighted average of the values from each model family and region fit. For the FIRE data, we determine the RV, $v \sin i$, and $\log g$ from our narrow-region fits (Table~\ref{table:modelfits}), and the $T_{\rm eff}$ and $f_{\rm sed}$ from our full-band fits (Table~\ref{table:modelfitsbroad}). For the GNIRS data we determine all parameters from the full wavelength coverage available.
We assign the weights in the weighted average as $e^{-\chi^2_{\rm reduced}}$, where the $\chi^2_{\rm reduced}$ for each best-fit model is given in Tables~\ref{table:modelfits} and~\ref{table:modelfitsbroad}. We report the final values from the weighted averages in Table~\ref{table:inclinations}, with the unbiased weighted sample standard deviation as our uncertainties.

We describe the outcomes of our model fitting and $\chi^2$ analysis in detail for each target in Sections~\ref{sec:0348models}--\ref{sec:0407models}.

\begin{deluxetable}{l l c c c c c c c } 
\tabletypesize{\footnotesize}
\tablecolumns{9} 
\tablewidth{0pt}
\tablecaption{Best-fit Photospheric Model Parameters for the Narrow-wavelength Regions  \label{table:modelfits} } 
\tablehead{
\colhead{Model} & \colhead{Region} & \colhead{Wavelength} & \colhead{$T_{\rm eff}$} & \colhead{$f_{\rm sed}$} & \colhead{$\log g$} &  \colhead{$v\sin i$} & \colhead{RV} & \colhead{$\chi^2_{\rm reduced}$} \\
\colhead{} & \colhead{} & \colhead{($\mu$m)} & \colhead{(K)} & \colhead{} & \colhead{(dex)} & \colhead{(km\,s$^{-1}$)} & \colhead{(km\,s$^{-1}$)} & \colhead{}
}
\startdata 
\multicolumn{9}{c}{2MASS J03480772$-$6022270 (T7, FIRE data)}\\
BT-Settl & J narrow & 1.260 -- 1.300 & $950 \pm 25$  & \nodata 	& $5.50 \pm 0.25$ & $102.4	\pm 3.9$ & $-11.8 \pm 0.8$ & 1.2 \\
BT-Settl & H narrow & 1.520 -- 1.562 & $950 \pm 25$  & \nodata 	& $5.00 \pm 0.25$ & $94.9 \pm 1.5$ & $-14.1 \pm 0.9$ & 0.9 \\
BT-Settl & K narrow & 2.110 -- 2.190 & $700 \pm 25$  & \nodata 	& $4.50 \pm 0.25$ & $115.4 \pm 2.2$ & $-17.1 \pm 1.3$ & 2.0 \\
Morley   & J narrow & 1.260 -- 1.300 & $900 \pm 50$  & 4        & $5.50 \pm 0.25$ & $ 105.7 \pm 1.8 $ & $-15.1 \pm 0.9$ & 1.6 \\
Morley   & H narrow & 1.520 -- 1.562 & $1000 \pm 50$ & 5        & $5.50 \pm 0.25$ & $ 114.3 \pm 2.2 $ & $-18.0 \pm 1.0$ & 2.1 \\	
Morley   & K narrow & 2.110 -- 2.190 & $800 \pm 50$  & 5        & $5.00 \pm 0.25$ & $ 103.2 \pm 1.9 $ & $-16.5 \pm 1.3$ & 2.1 \\
Sonora   & J narrow & 1.260 -- 1.300 & $1000 \pm 50$ & \nodata  & $5.00 \pm 0.13$ & $ 96.5 \pm 1.5$  & $-12.6 \pm 0.8$ & 1.6 \\
Sonora   & H narrow & 1.520 -- 1.562 & $1000 \pm 50$ & \nodata  & $5.00 \pm 0.13$ & $ 110.7 \pm 1.5$ & $-14.2 \pm 0.9$ & 1.1 \\	
Sonora   & K narrow & 2.110 -- 2.190 & $800 \pm 50$  & \nodata  & $4.75 \pm 0.13$ & $ 99.6 \pm 2.6$ &	$-11.2 \pm 1.4$ & 2.0 \\
\multicolumn{2}{l}{Adopted values} & \nodata & \nodata  & \nodata  & $5.1 \pm 0.3$ & $103.5 \pm 7.4$ &	$-14.1 \pm 3.7$ & \nodata \\
\hline
\multicolumn{9}{c}{2MASS J12195156$+$3128497 (L8, FIRE data)}\\
BT-Settl	     &	H narrow 1	& 1.500 -- 1.550 & $1250 \pm 25$ & \nodata  & $5.00 \pm 0.25$ &	$77.4 \pm 2.6$ & $-17.2	\pm 1.6$ & 1.3 \\
BT-Settl	     &	H narrow 2	& 1.720 -- 1.780 & $1150 \pm 25$ & \nodata  & $4.00 \pm 0.25$ &	$85.7 \pm 1.4$ & $-19.0	\pm 0.9$ & 2.6 \\
BT-Settl	     &	K narrow	& 1.970 -- 2.055 & $1400 \pm 25$ & \nodata  & $5.00 \pm 0.25$ &	$76.8 \pm 1.4$ & $-16.6 \pm 1.1$ & 2.7 \\
SM08            &	H narrow 1	& 1.500 -- 1.550 & $1400 \pm 50$ & 4        & $5.50 \pm 0.25$ & $78.1 \pm 2.4$ & $-19.6 \pm 1.4$ & 1.4 \\
SM08            &	H narrow 2  & 1.720 -- 1.780 & $1500 \pm 50$ & 4        & $5.00 \pm 0.25$ & $84.3 \pm 1.3$ & $-25.9 \pm 0.9$ & 2.6 \\
SM08            &	K narrow   & 1.970 -- 2.055  & $1400 \pm 50$ & 2        & $5.50 \pm 0.25$ & $77.1 \pm 1.5$ & $-20.0 \pm 1.1$ & 2.7 \\
\multicolumn{2}{l}{Adopted values} & \nodata & \nodata & \nodata  & $5.1 \pm 0.5$ & $79.0 \pm 3.4$ &	$-19.0 \pm 4.2$ & \nodata \\
\hline
\multicolumn{9}{c}{2MASS J04070752$+$1546457 (L3.5, GNIRS data)}\\
BT-Settl & K & 2.275 -- 2.332 & $1700 \pm 25$ & \nodata  & $5.00 \pm 0.25$ & $82.7 \pm 0.9$ & $43.7 \pm 0.9$ & 1.0 \\
SM08 & K & 2.275 -- 2.332     & $2000 \pm 50$ & 4        & $5.50 \pm 0.25$ & $ 82.4 \pm 0.9$ & $43.1 \pm 0.8$ & 1.1 \\
\multicolumn{2}{l}{Adopted values} & \nodata & $1840 \pm 210$ & 4  & $5.2 \pm 0.4$ & $82.6 \pm 0.2$ &	$43.4 \pm 2.1$ & \nodata \\
\enddata
\tablecomments{Best-fit photospheric model parameters for our three L and T dwarfs over the narrow regions within each FIRE band, and over the entirety of the GNIRS spectrum. We fit each wavelength region independently. 
We adopt the $\log{g}$, $v \sin i$, and RV values determined from the narrow wavelength regions of the FIRE spectra of the T7 and L8 dwarfs. The $T_{\rm eff}$ and $f_{\rm sed}$ estimates are adopted from the full-band fits (Table \ref{table:modelfitsbroad}), although we include the findings $T_{\rm eff}$ and $f_{\rm sed}$ from the narrow region fitting for completeness. 
For the GNIRS data of the L3.5 dwarf we adopt all parameters from the wavelength region shown here. The $f_{\rm sed}$ parameter is only applicable to the SM08 and Morley models. The adopted values are the weighted averages for each object, where the weights are $e^{-\chi^2_{\rm reduced}}$, and the uncertainties are the unbiased weighted sample standard deviations (as described in Section~\ref{sec:SpectraAnalysis}). The adopted RVs include systematic uncertainties of $\pm$3.0~km~s$^{-1}$ (for the T7 and L8 dwarfs) or $\pm$2.0~km~s$^{-1}$ (for the L3.5 dwarf) added in quadrature to account for the wavelength calibration uncertainties of the FIRE and GNIRS spectra, respectively (Sec.~\ref{sec:spectra}).
}
\end{deluxetable}

\begin{deluxetable}{l l c c c c c c c} 
\tablecolumns{9} 
\tablewidth{0pt}
\tablecaption{Best-fit Photospheric Model Parameters for the Full Bands \label{table:modelfitsbroad} } 
\tablehead{
\colhead{Model} & \colhead{Band} & \colhead{Wavelength} & \colhead{$T_{\rm eff}$} & \colhead{$f_{\rm sed}$} & \colhead{$\log{g}$\tablenotemark{a}} & \colhead{$v\sin{i}$\tablenotemark{a}} & \colhead{RV\tablenotemark{a}} & \colhead{$\chi^2_{\rm reduced}$} \\
\colhead{} & \colhead{} & \colhead{($\mu$m)} & \colhead{(K)} & \colhead{} & \colhead{(dex)} & \colhead{(km\,s$^{-1}$)} & \colhead{(km\,s$^{-1}$)} & \colhead{}
}
\startdata 
\multicolumn{9}{c}{2MASS J03480772$-$6022270 (T7, FIRE data)}\\
BT-Settl & J & 1.140 -- 1.345 & $900 \pm 25$   & \nodata  & $5.0 \pm 0.25$  & 118.3 & -14.2 & 7.4 \\
BT-Settl & H & 1.480 -- 1.790 & $750 \pm 25$   & \nodata  & $4.5 \pm 0.25$  & 118.3 & -11.3 & 11 \\
BT-Settl & K & 1.960 -- 2.350 & $700 \pm 25 $  & \nodata  & $4.0 \pm 0.25$  & 107.9 & -13.7 & 2.8 \\
Morley   & J & 1.140 -- 1.345 & $800 \pm 50 $  & 5        & $4.0 \pm 0.25$  & 118.3 & -15.1 & 10 \\
Morley   & H & 1.480 -- 1.790 & $700 \pm 50 $  & 5        & $4.0 \pm 0.25$  & 118.3 & -10.0 & 28 \\	
Morley   & K & 1.960 -- 2.350 & $900 \pm 50 $  & 5        & $4.0 \pm 0.25$  & 107.2 & -19.5 & 2.5 \\
Sonora   & J & 1.140 -- 1.345 & $900 \pm 50 $  & \nodata  & $5.25 \pm 0.13$ & 94.0 & -16.8 & 4.2 \\
Sonora   & H & 1.480 -- 1.790 & $850 \pm 50 $  & \nodata  & $4.25 \pm 0.13$ & 118.3 & -21.5 & 2.0 \\	
Sonora   & K & 1.960 -- 2.350 & $1000 \pm 50 $ & \nodata  & $4.50 \pm 0.13$ & 104.5 & -15.5 & 2.3 \\
\multicolumn{2}{l}{Adopted values} & \nodata & $880 \pm 110$ & 5  & \nodata & \nodata & \nodata & \nodata \\
\hline
\multicolumn{9}{c}{2MASS J12195156$+$3128497 (L8, FIRE data)}\\
BT-Settl	     &	J	& 1.140 -- 1.345 & $1400 \pm 25$ & \nodata  & $5.50 \pm 0.25$ & 72.2 & -24.5 & 2.3 \\
BT-Settl	     &	H	& 1.480 -- 1.790 & $1200 \pm 25$ & \nodata  & $4.50 \pm 0.25$ & 85.8 & -18.9 & 2.0 \\
BT-Settl	     &	K	& 1.960 -- 2.350 & $1250 \pm 25$ & \nodata  & $4.50 \pm 0.25$ & 85.8 & -17.8 & 2.1 \\
SM08             &	J   & 1.140 -- 1.345 & $1200 \pm 50$ & 3        & $5.50 \pm 0.25$ & 72.2 & -14.6 & 2.0 \\
SM08             &	H   & 1.480 -- 1.790 & $1500 \pm 50$ & 3        & $5.00 \pm 0.25$ & 83.2 & -23.2 & 1.9 \\
SM08             &	K   & 1.960 -- 2.350 & $1500 \pm 50$ & 3        & $4.50 \pm 0.25$ & 85.8 & -18.6 & 2.4 \\
\multicolumn{2}{l}{Adopted values} & \nodata & $1330 \pm 140$ & 3 & \nodata & \nodata & \nodata & \nodata \\
\enddata
\tablecomments{Best-fit photospheric model parameters for the FIRE data of the T7 and L8 dwarfs, fit over each of the full $J$, $H$, and $K$ bands. We use these fits to inform our final $T_{\rm eff}$ and $f_{\rm sed}$ determinations. The adopted values are the weighted averages (as described in Section~\ref{sec:SpectraAnalysis}). 
The GNIRS data for the L3.5 dwarf are not shown here as they only cover a narrow-wavelength region, and so the fitting results for that object are shown in their entirety in Table~\ref{table:modelfits}. 
\tablenotetext{a}{ 
We adopt the $\log g$, RV and $v \sin i$ values from the narrow-wavelength regions (Table~\ref{table:modelfits}).  To better determine $T_{\rm eff}$ and $f_{\rm sed}$, we still allow $\log g$ to vary unconstrained in the full-band fitting, while RV and $v\sin i$ are allowed to probe within $2\sigma$ of the adopted values. In some cases the best-fitting RVs and $v\sin i$ values correspond to the extremes of their allowed range, so we do not include uncertainties for them here.
} 
} 
\end{deluxetable}

\subsection{2MASS J0348$-$6022 (T7)}
\label{sec:0348models}

\begin{figure} 
  \centering
    \includegraphics[trim={0.5cm 6.2cm 1.cm 2.5cm},clip,width=0.95\textwidth]{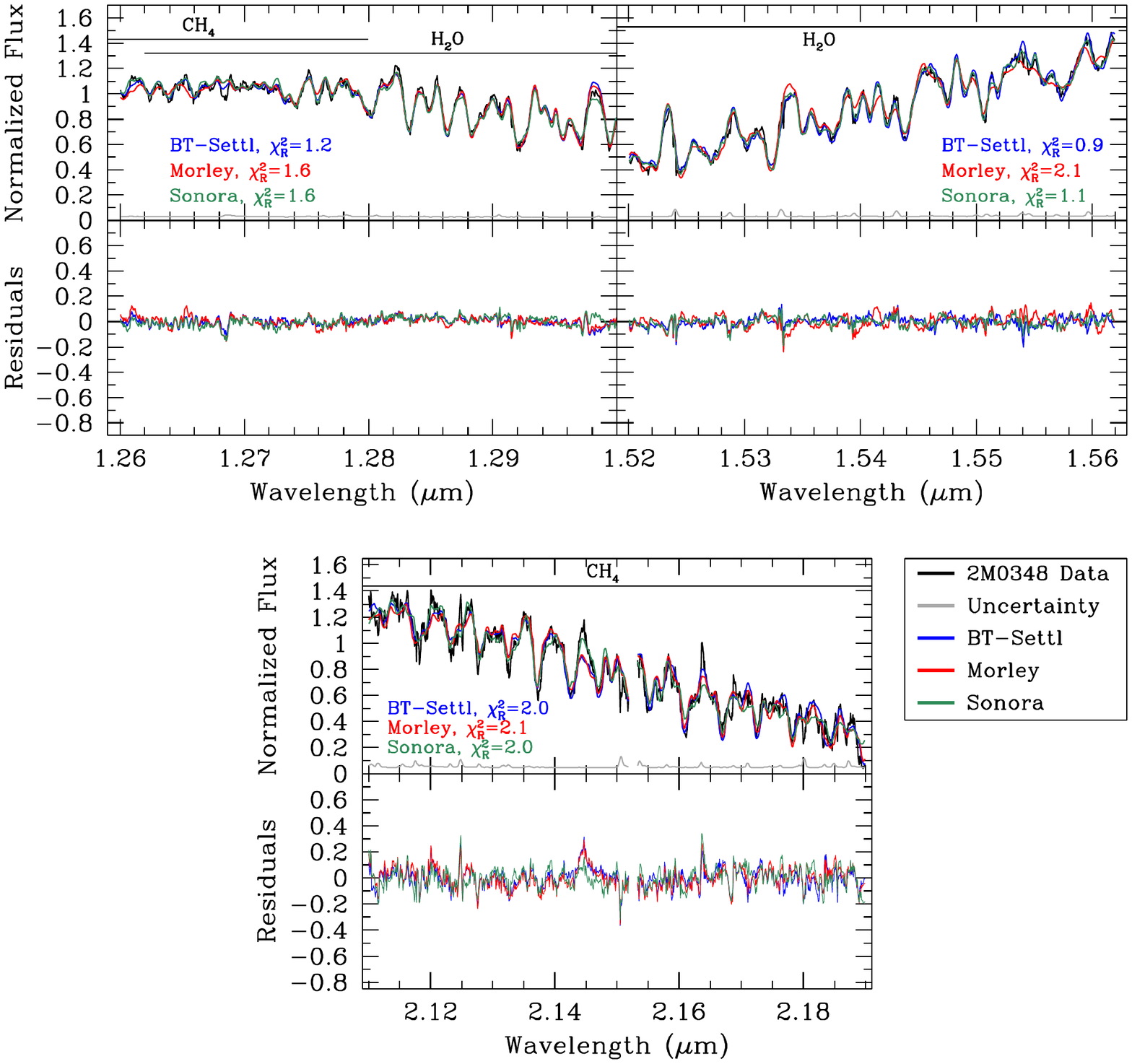}
  \caption{Three narrow regions within the $J$ (top left), $H$ (top right), and $K$ (bottom) bands of our $R\approx6000$ FIRE spectra of 2MASS J0348$-$6022 (T7), with the best-fitting BT-Settl, Morley, and Sonora models. The parameters of the best-fit models shown here are listed in Table~\ref{table:modelfits}. These regions were selected for their density of H$_2$O and CH$_4$ absorption lines \citep{BDSS2003, cushing_etal06, Bochanski2011} to allow precise RV, $v\sin i$, and $\log g$ determinations. The reduced $\chi^2$ statistic is shown for each model ($\chi^2_R$). The residuals are shown in the lower section of each panel, where the colors match those of the corresponding models. 
} 
  \label{fig:0348spectrum_narrow}
\end{figure} 


\begin{figure}
  \centering
    \includegraphics[trim={0.5cm 6.2cm 1.cm 2.5cm},clip,width=0.95\textwidth]{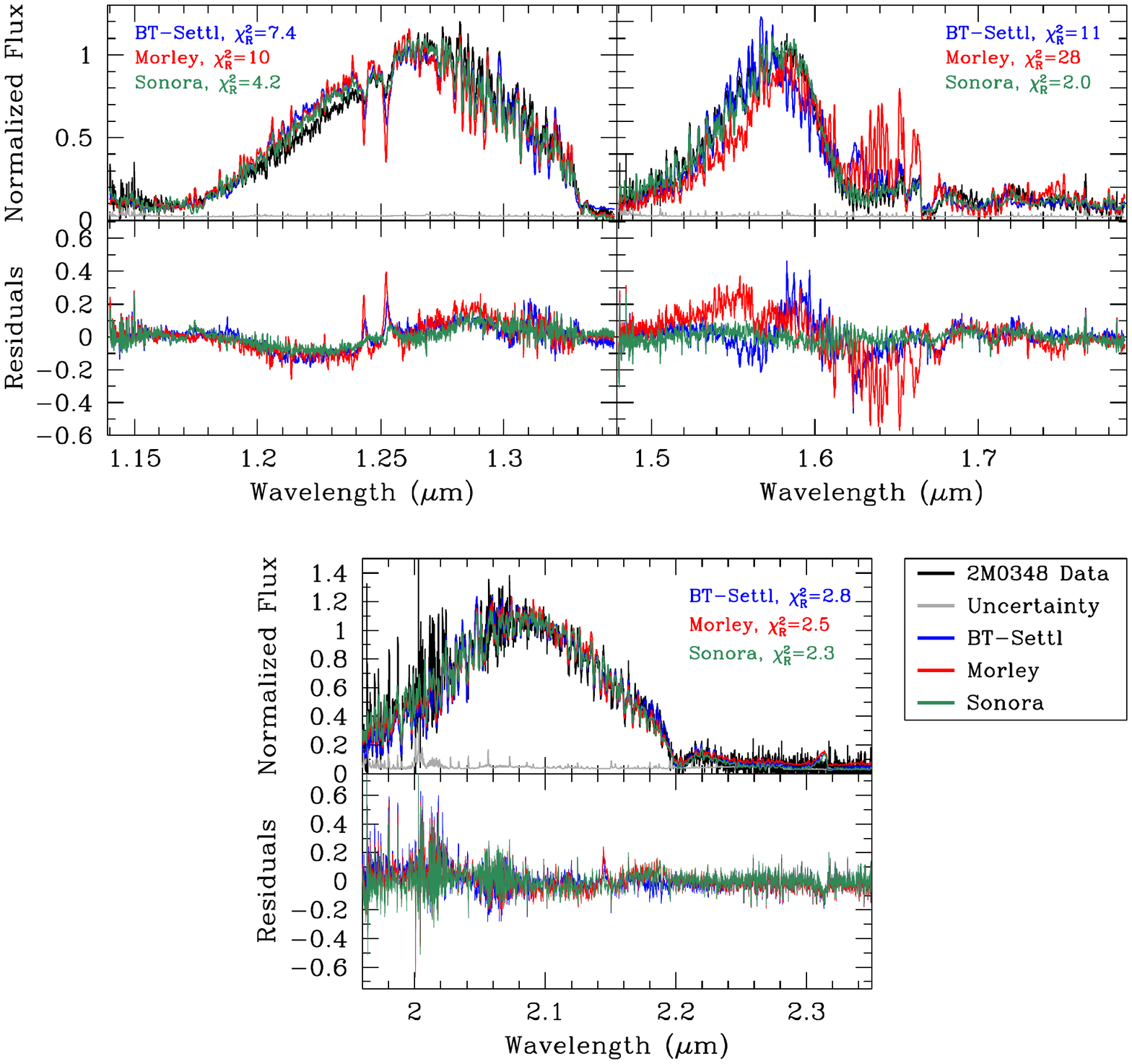}
  \caption{FIRE $R\approx6000$ $J$- (top left), $H$- (top right), and $K$-band (bottom) spectra of 2MASS J0348$-$6022 (T7) with the best-fitting BT-Settl, Morley, and Sonora models for each band. The parameters of the best-fit models shown here are listed in Table~\ref{table:modelfitsbroad}. The reduced $\chi^2$ statistic is shown for each model. The residuals are shown in the lower sections of each panel, where the colors match those of the corresponding models. 
} 
  \label{fig:0348spectrum_full}
\end{figure}

Based on its T7 spectral type, 2MASS J0348$-$6022 is expected to have an effective temperature $T_{\rm eff} \lesssim 1000$~K (e.g., \citealt{Stephens2009, Filippazzo2015}). Its photosphere should be governed by gas opacity, with a cloud layer buried deeply \citep[$f_{\rm sed}\geq3$;][]{Ackerman2001, Marley2002} within the atmosphere. 
Thus we expect the atmosphere of this object to be relatively clear and cloudless. So, the cloud-free Sonora models are appropriate for fitting this object's spectra.
A cloudless atmosphere does not imply a completely homogeneous surface, and it is possible that one of the alternative mechanisms presented in Section~\ref{sec:SpitzerDiscussion} (e.g., temperature variations; \citealt{Robinson2014}) is responsible for the observed variability. 
We also compared this target to the BT-Settl and Morley models. The effective temperature grid of the available SM08 models does not extend to the low temperatures expected for a T7 spectral type. 

We selected a grid of parameters ranging from $T_{\rm eff} = 700$ to 1000~K, $\log g = 4.0$ to 5.5~dex in steps of 0.5~dex (0.25 for the Sonora models), and, for the Morley models, condensate sedimentation efficiencies from $f_{\rm sed} = 2$ to 5 in unit steps. We selected our $\log g$ grid based on the range in surface gravities predicted by the SM08 evolutionary models for brown dwarfs.  We selected the RV and $v \sin i$ grids by first testing a wide, coarse grid to determine  approximate RV and $v \sin i$ values.  We then narrowed it down to between $-$5\,km\,s$^{-1}$ and $-$30\,km\,s$^{-1}$ for RV and between 75\,km\,s$^{-1}$ and 115\,km\,s$^{-1}$ for $v \sin i$, both in steps of 0.1\,km\,s$^{-1}$.

We find that a wide range in parameters fit the $z$ band equally well, and it is therefore not diagnostic for our study. We exclude the $z$ band from our analysis, and only consider the $J-$, $H-$, and $K-$ band spectra for this target. To reliably determine $\log g$, RV, and $v \sin i$, we selected narrow regions dominated by molecular lines within each band: the 1.26--1.30 $\mu$m $J$-band region dominated by H$_2$O and CH$_4$ (Fig.~\ref{fig:0348spectrum_narrow}, top left), the 1.520--1.562 $\mu$m $H$-band region dominated by H$_2$O (Fig.~\ref{fig:0348spectrum_narrow}, top right), and the 2.11-2.19 $\mu$m $K$-band region containing primarily CH$_4$ lines (Fig.~\ref{fig:0348spectrum_narrow}, bottom). 
The best-fit photospheric models for the narrow wavelength regions are shown in Figure~\ref{fig:0348spectrum_narrow} and for the full bands in Figure~\ref{fig:0348spectrum_full}. The high $f_{\rm sed}$ values of the \citet{Morley2012} models in all of the full band fits indicate an optically thin, relatively cloudless atmosphere, as expected for a late-T type brown dwarf.

We adopt the weighted average and unbiased weighted sample standard deviation (Sec.~\ref{sec:SpectraAnalysis}) of the values in Table~\ref{table:modelfits} as our estimates for $\log g$, $v \sin i$, and RV. For $v \sin i$ in particular, we find a very high degree of rotational broadening: $v \sin i = 103.5 \pm 7.4$\,km\,s$^{-1}$. This is consistent with the short rotational period (Sec.~\ref{sec:SpitzerDiscussion}), and will be discussed further in Section~\ref{sec:Discussion}. The adopted parameters from the spectroscopic fitting are shown in Table~\ref{table:inclinations}.

\subsection{2MASS J1219+3128 (L8)}
\label{sec:1219models}
 
\begin{figure}
  \centering
    \includegraphics[trim={0.5cm 6.2cm 1.cm 2.5cm},clip,width=0.95\textwidth]{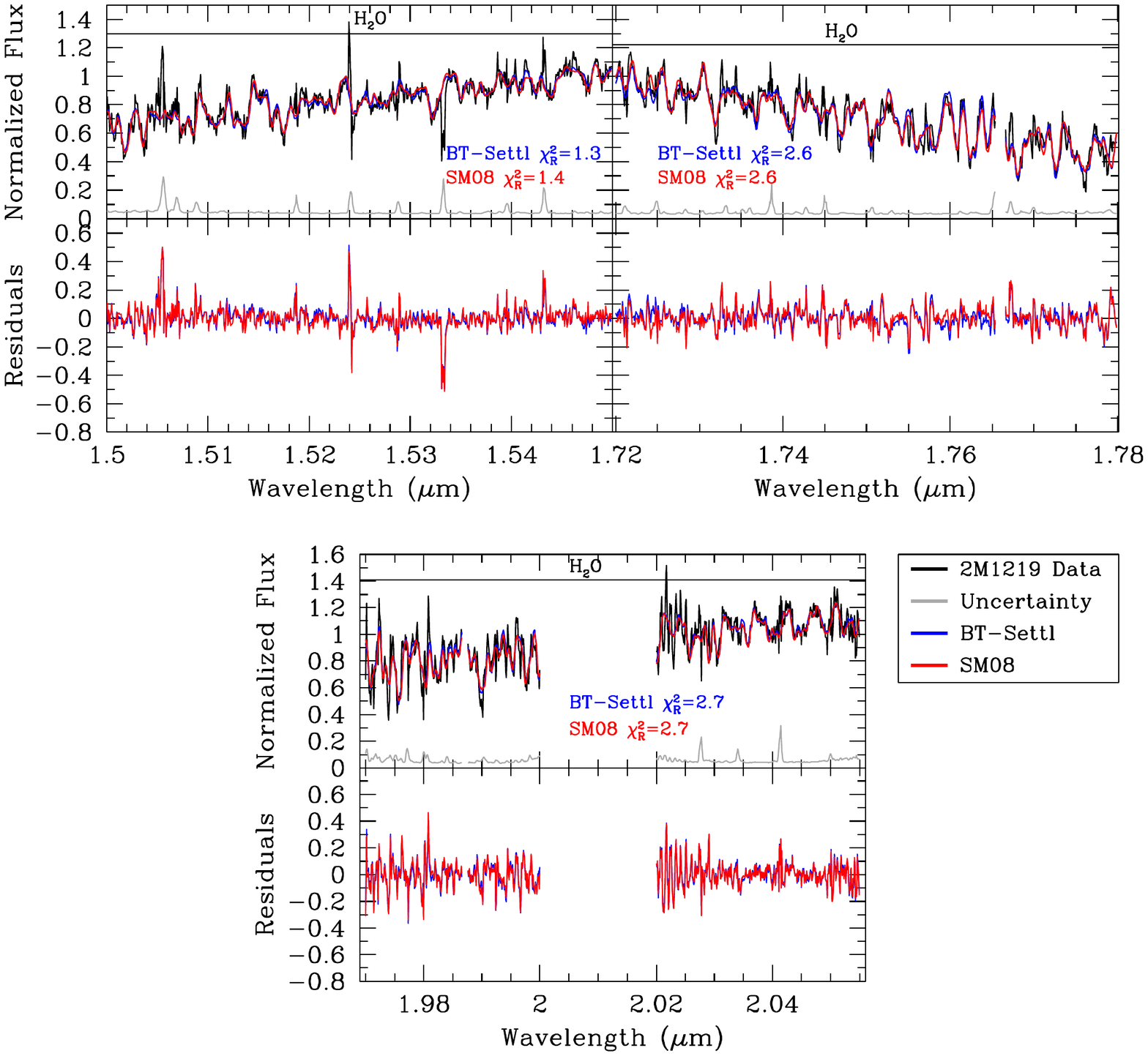}
  \caption{Three narrow regions within the $H$ (top left and right) and $K$ (bottom) bands of our $R\approx6000$ FIRE spectrum of 2MASS J1219$+$3128 (L8), with the best-fitting BT-Settl, and SM08 models. These regions were selected for their density of H$_2$O absorption lines \citep{BDSS2003, cushing_etal06} to allow precise $v\sin i$, RV, and $\log g$ determinations. The parameters of the models shown here are listed in Table~\ref{table:modelfits}. The reduced $\chi^2$ statistic is shown for each model. The residuals are shown in the lower sections of each panel, where the colors match those of the corresponding models. 
  A strong telluric feature has been masked between 2.00 and 2.02 $\mu$m in the $K$ band.
 } 
  \label{fig:1219spectrum_narrow}
\end{figure}

\begin{figure}
  \centering
    \includegraphics[trim={0.5cm 6.2cm 1.cm 2.5cm},clip,width=0.95\textwidth]{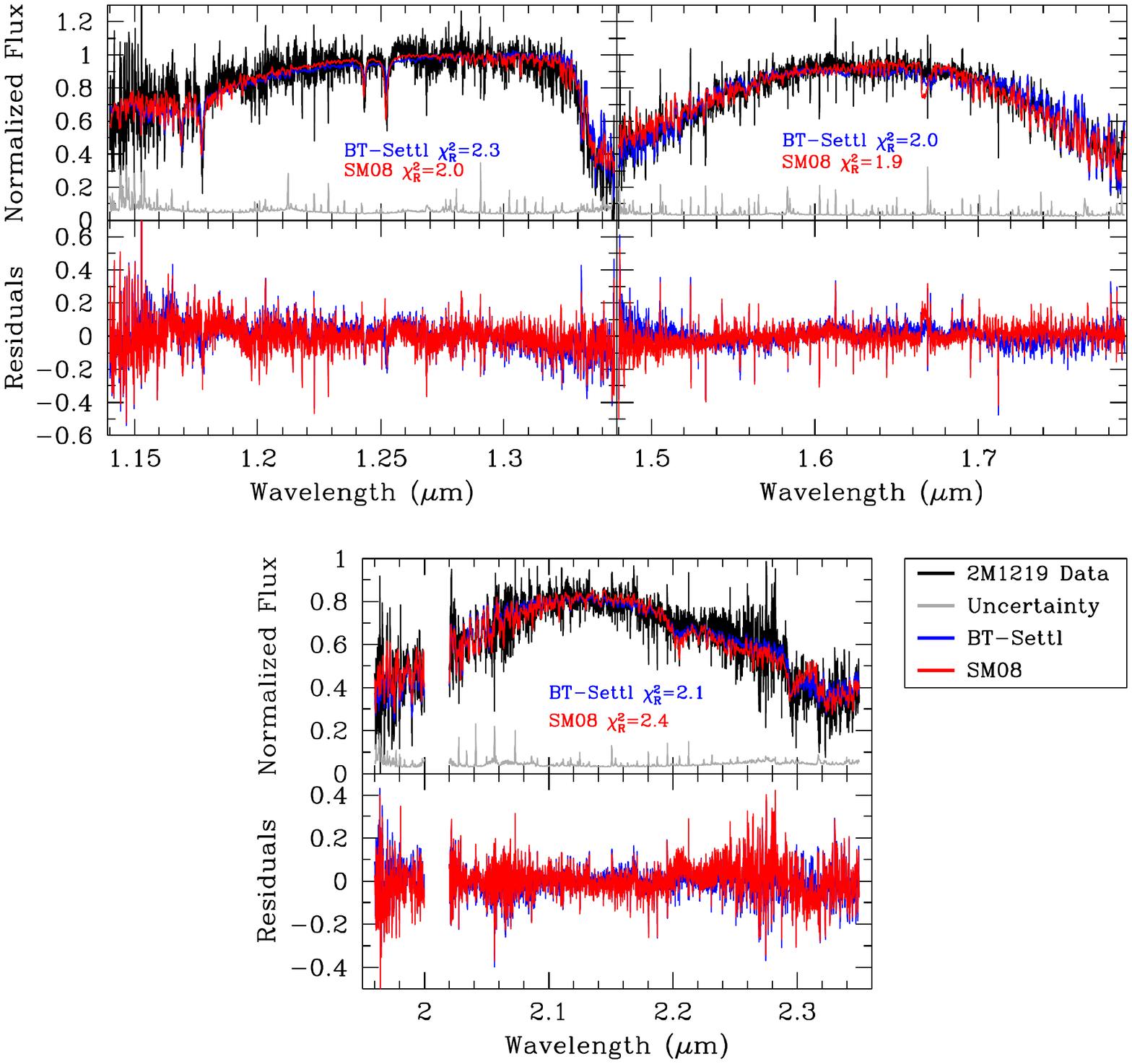}
  \caption{FIRE $R\approx6000$ $J$- (top left), $H$- (top right), and $K$-band (bottom) spectra of 2MASS 1219$+$3128 (L8) with the best-fitting BT-Settl and SM08 models for the entire bands. The parameters of the models shown here are listed in Table~\ref{table:modelfitsbroad}. The reduced $\chi^2$ statistic is shown for each model. The residuals are shown in the lower sections of each panel, where the colors match those of the corresponding models. The best-fitting SM08 model has $T_{\rm eff}=1500$~K and $\log g=5.0$ and shows a methane feature at 1.665~$\mu$m. Despite the presence of this feature in the model and its absence in the data, the shown photospheric model offers the best overall fit to the $H$-band spectrum. The appearance of the methane feature in the photospheric model may suggest that this L8 dwarf is close to transitioning to a T-type atmosphere. A strong telluric feature has been masked between 2.00 and 2.02 $\mu$m in the $K$ band.
} 
  \label{fig:1219spectrum_full}
\end{figure}

Based on its L8 spectral type, we expect 2MASS J1219+3128 to have an effective temperature of $T_{\rm eff} \sim 1400$~K (e.g., \citealt{Stephens2009, Filippazzo2015}). The Morley models are not suitable for this target, as that model grid extends to a maximum of $T_{\rm eff}$ = 1300 K. The Sonora models are also not appropriate, as they are cloud-free, while late-L dwarfs are very dusty and are expected to have thick, patchy clouds. Therefore, we instead adopt only the SM08 and BT-Settl models as they cover sufficiently high temperatures for this spectral type and include treatments of dust. We selected a grid of parameters ranging from $T_{\rm eff}$ = 1100~K to 1700~K, $\log g$ = 4.0 to 5.5~dex, and for the SM08 models, condensate sedimentation efficiencies from $f_{\rm sed}$ = 1 to 4 in unit steps. We selected the RV and $v \sin i$ grids using the same method as before (Sec.~\ref{sec:0348models}): by first testing a large, coarse grid to determine the approximate RV and $v \sin i$ values. The final grid was between $-$5\,km\,s$^{-1}$ and $-$30\,km\,s$^{-1}$ for RV and between 70\,km\,s$^{-1}$ and 110\,km\,s$^{-1}$ for $v \sin i$, both in steps of 0.1\,km\,s$^{-1}$.

As for 2MASS~J0348--6022 (Sec.~\ref{sec:0348models}), we find that a wide range in parameters fit the $z$ band equally well.  It is not diagnostic for our study, and we exclude the $z$ band from our analysis. The $J$-band data had fairly low signal-to-noise ratio and had no regions with clearly defined lines from which we could measure $v \sin i$. We instead selected two narrow regions in the $H$ band, along with a narrow region in the $K$ band. In the $H$ band we selected 1.50--1.55~$\mu$m and 1.72--1.78~$\mu$m, where the first region is dominated by H$_2$O, and the second is dominated by FeH, H$_2$O, and potentially some CH$_4$. 
The best lines in our data set for measuring $v \sin i$ in the $K$ band are H$_2$O lines between 1.970~$\mu$m and 2.055~$\mu$m, located on either side of a major telluric feature where our data have very low quality. We opted to mask out this region (2.00--2.02~$\mu$m) before fitting the models. We show the narrow band fits in Figure~\ref{fig:1219spectrum_narrow}, and the full-band fits in Figure~\ref{fig:1219spectrum_full}.

We also find a high degree of rotational broadening for 2MASS J1219+3128, with $v \sin i = 79.0 \pm 3.4$ km\,s$^{-1}$ (Table~\ref{table:modelfits}). This is consistent with the short photometric period for this object. The adopted parameters from the spectroscopic fitting are shown in Table~\ref{table:inclinations}.

\subsection{2MASS J0407+1546 (L3.5)}
\label{sec:0407models} 
 
\begin{figure}
  \centering
    \includegraphics[trim={1.7cm 12.5cm 0.5cm 2.9cm},clip,width=0.9\textwidth]{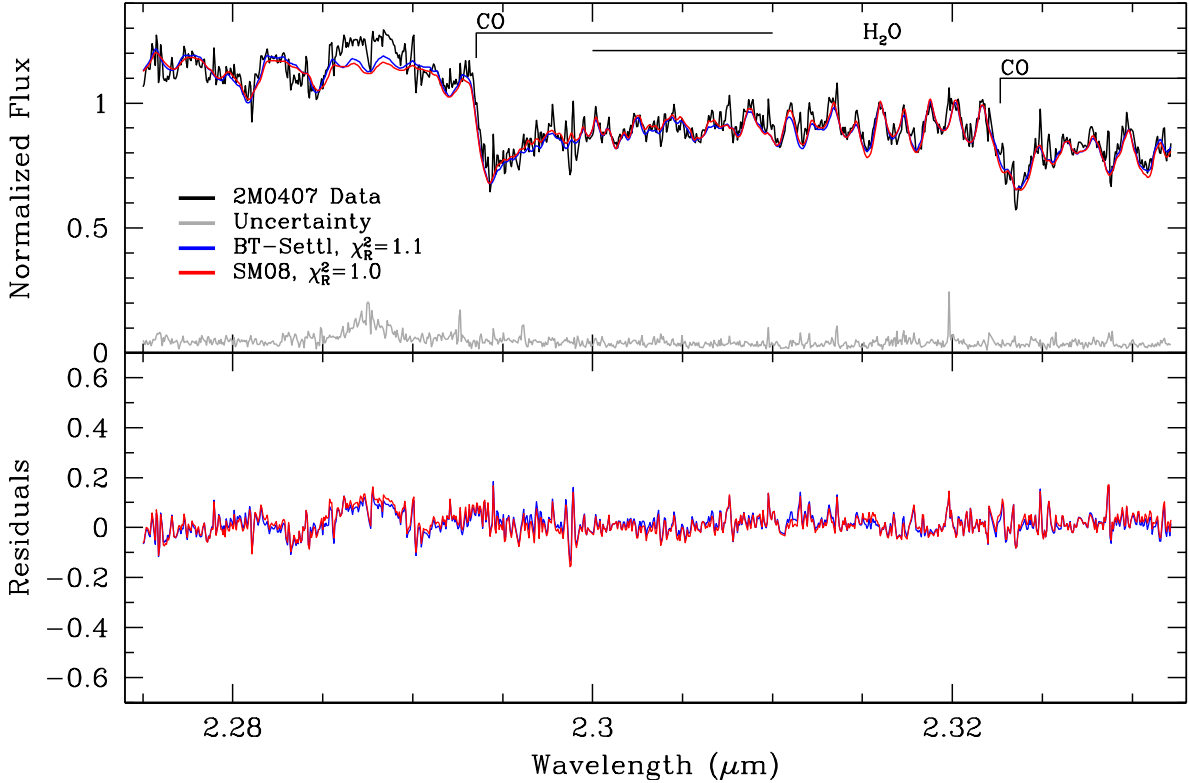}
  \caption{GNIRS 2.275--2.332 $\mu$m $R\approx12,000$ spectrum of 2MASS J0407$+$1546, with the best-fitting SM08 and BT-Settl models overlaid. The parameters of the models shown here are listed in Table~\ref{table:modelfits}. The reduced $\chi^2$ statistic is shown for each model. Major molecular features \citep{cushing_etal06} are indicated. The residuals are shown in the lower panel, where the colors match those of the corresponding models.
 } 
  \label{fig:0407spectrum}
\end{figure} 

Based on its L3.5 spectral type, 2MASS J0407+1546 is expected to have an effective temperature of $\sim$1800~K \citep[e.g.,][]{Stephens2009, Filippazzo2015}, with a fairly cloudy atmosphere. We therefore select the SM08 and BT-Settl models. We do not include the Sonora models as they are cloudless, or the Morley models as they are for temperatures below 1300~K.
We selected the following parameter grid for fitting: $T_{\rm eff}$ = 1500~K to 2100~K in steps of 100~K, $\log g$ = 4.0 dex to 5.5~dex in steps of 0.5~dex, and condensate sedimentation efficiency $f_{\rm sed} = 1$ to 4 in unit steps. We selected 30\,km\,s$^{-1}$ to 60\,km\,s$^{-1}$ for RV and 75\,km\,s$^{-1}$ to 100\,km\,s$^{-1}$ for $v \sin i$, both in steps of 0.1\,km\,s$^{-1}$.  
Our GNIRS observations cover the narrow region from 2.275 $\mu$m to 2.332~$\mu$m, which contains primarily H$_2$O and CO lines.  We show the best-fitting models in Figure~\ref{fig:0407spectrum}. 

The narrower-wavelength coverage of our GNIRS data means we have limited effective temperature and sedimentation efficiency information compared to the full-band spectra of the two other objects. Although we cannot place a high confidence on the results for these two parameters, we find that the effective temperature is consistent with expectations for an L3.5 dwarf, with $T_{\rm eff} = 1840 \pm 210$~K. We find a high degree of rotational broadening, at $v \sin i = 82.6 \pm 0.2$~km\,s$^{-1}$, consistent with the short rotational period. Table~\ref{table:inclinations} lists all of the physical parameters detemined from the spectroscopic fits.

\section{Discussion}
\label{sec:Discussion}

\subsection{The Three Most Rapidly Rotating Ultra-cool Dwarfs: Possibility for Auroral Emissions}

\begin{figure}
  \centering
    \includegraphics[trim={1.1cm 6.1cm 1.75cm 3.3cm},clip,width=0.7\textwidth]{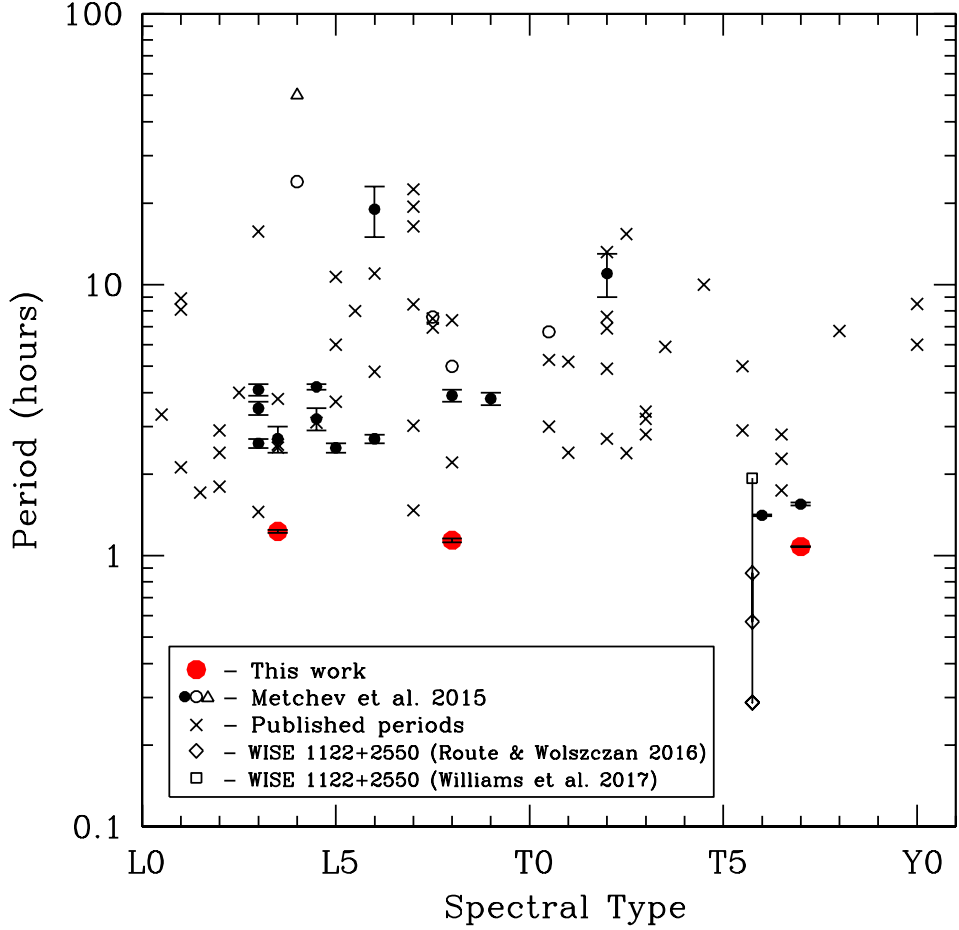}
  \caption{
  Rotation period as a function of spectral type for all 78 periodically variable L0--Y0 dwarfs known as of this writing. The full list of rotation periods is given in the Appendix in Table~\ref{table:allperiods} with references. 
  The ``ultra-fast rotators'' of this work are shown in red. Black circles are periods from \citet{Metchev2015}, where solid circles denote variables with well-determined periodicities and open circles denote variables with uncertainties of $\geq 50\%$. An upward-facing triangle denotes the 50 hour lower limit on the periodicity of 2MASS 1753$-$6559. Open diamonds denote the possible period harmonics of WISEPC J1122+2550 \citep{Route2016}, while an open square denotes the revised rotational period for this target from \citet{Williams2017}. The points for WISEPC J1122+2550 are offset slightly to the left to avoid ambiguity with another T6 dwarf. Other previously published periods are denoted by the ``$\times$'' symbol. 
  }
  \label{fig:periodicity} 
\end{figure}

The ${1.080}^{+0.004}_{-0.005}$~h, ${1.14}^{+0.03}_{-0.01}$~h, and ${1.23}^{+0.01}_{-0.01}$~h photometric periods of our three L and T dwarfs are shorter than any others yet observed (Fig.~\ref{fig:periodicity}; Table~\ref{table:allperiods}). The previously reported shortest photometric period for an ultra-cool dwarf was $1.41\pm0.01$ h for the T6 2MASS J22282889$-$4310262 \citep{Clarke2008, Buenzli2012, Metchev2015}. 
\citet{Route2016} have reported an even shorter possible period, 0.288 h for the T6 dwarf WISEPC J112254.73+255021.5, based on radio flare observations from the Arecibo Observatory radio telescope. However, they indicate that the 0.288 h period may be a harmonic of a longer period, or that the flares may in fact not have been periodic. They base their period on five flaring events with the first and last separated by $\sim $240 days. When analyzing the data with the flares removed, they do not find any indication of variability. A later study by \citet{Williams2017} using the Very Large Array confirmed the same object to have variable polarized emission, but with a longer period of 1.93 h. They also observed the target photometrically in the $z$ band using Gemini/GMOS-N, and did not find any indication of variability. We therefore do not consider WISEPC J112254.73+255021.5 as an ultra-fast rotator, leaving the three objects reported here as the fastest known L or T dwarf rotators.  

We find a high degree of rotational line broadening for all three targets, consistent with the short photometric periods. At projected rotation velocities of 103.5 km\,s$^{-1}$ for 2MASS~J0348$-$6022 (T7), 79.0 km\,s$^{-1}$ for 2MASS~J1219$+$3128 (L8) and 82.6 km\,s$^{-1}$ for 2MASS~J0407$+$1546 (L3.5), these are among the most rapidly spinning ultra-cool dwarfs known to date. In the comprehensive compilation of ultra-cool dwarf rotation measurements by \citet{Crossfield2014}, he lists only two other ultra-cool dwarfs with $v\sin i > 80$ km\,s$^{-1}$: HD~130948C (86 km\,s$^{-1}$, L4) and LP~349--45B (83 km\,s$^{-1}$, M9), both from \citet{Konopacky2012}. 

The rapid projected rotational velocities of our targets confirm that the $\sim 1$ h periodicities of their light curves correspond to their true rotation periods, and that they are not more slowly rotating brown dwarfs with multiple spots at semi-regular longitudinal intervals, as seen on Jupiter \citep{dePater2016}, or beat patterns arising from planetary-scale waves \citep{Apai2017}.

\begin{deluxetable}{l c c c} 
\tablecolumns{4} 
\tablewidth{0pt}
\tablecaption{Physical Parameters for the Three L and T Dwarfs \label{table:inclinations} }
\tablehead{
\colhead{Parameter} & \colhead{2MASS J0348-6022} & \colhead{2MASS J1219+3128} & \colhead{2MASS J0407+1546}
}
\startdata 
Spectral Type & T7 & L8 & L3.5 \\
$P_{\rm rot}$ (h)          & $1.080^{+0.004}_{-0.005}$ & $1.14^{+0.03}_{-0.01}$    & $1.23^{+0.01}_{-0.01}$ \\
$T_{\rm eff}$ (K)           & $880 \pm 110$             & $1330 \pm 140$            & $1840 \pm 210$ \\
$\log{g}$                   & $5.1 \pm 0.3$             & $5.1 \pm 0.5$             & $5.2 \pm 0.4$ \\
$v \sin i$ (km\,s$^{-1}$)   & $103.5 \pm 7.4$           & $79.0 \pm 3.4$            & $82.6 \pm 0.2$ \\
RV (km\,s$^{-1}$)         & $-14.1 \pm 3.7$           & $-19.0 \pm 4.2$           & $43.4 \pm 2.1$ \\
$R$ ($R_\odot$)             & $0.093^{+0.016}_{-0.010}$ & $0.100^{+0.027}_{-0.013}$ & $0.100^{+0.024}_{-0.008}$ \\
$M$ ($M_\odot$)             & $0.041^{+0.021}_{-0.017}$ & $0.047^{+0.022}_{-0.025}$ & $0.064^{+0.009}_{-0.027}$ \\
Age (Gyr)                   & $3.5^{+11.5}_{-2.9}$      & $0.9^{+12.8}_{-0.8}$      & $0.8^{+11.2}_{-0.65}$ \\
$v_{\rm eq}$ (km\,s$^{-1}$) & $105^{+18}_{-12}$   & $107^{+29}_{-15}$   & $99^{+24}_{-8}$ \\
Inclination (\degree)       & $81^{+9}_{-27} $    & $47^{+9}_{-17}$     & $57^{+7}_{-21}$\\
Oblateness                  & 0.08 & 0.08 & 0.05\\
\enddata 
\tablecomments{$P_{\rm rot}$ is determined from our photometric data. $T_{\rm eff}$, $\log{g}$, $v \sin i$, and RV are determined from our spectra by comparing to model photospheres. 
$R$, $M$, and the ages are determined by interpolation of the $\log g$-$T_{\rm eff}$ grid in the evolutionary models of SM08. The equatorial velocities ($v_{\rm eq}$) and spin-axis inclinations ($i$) are computed using the aforementioned values.\\
The evolutionary model radii listed are assumed to be the equatorial radii.  With oblateness factors between 0.05 and 0.08, the difference between the polar and equatorial radii is 5\%--8\%.  In reality, the evolutionary models (which ignore rotation) likely produce ``mean'' radii that are in between the equatorial and the polar radii.  Hence, any difference between the ``mean'' and the equatorial radii would be 3\%--4\%. This would revise our estimates for the equatorial velocities up by $\sim$3\%--4\%, but such systematic offsets would still be $\sim$3 times smaller than the quoted uncertainties. The effect on the inclination estimates would be negligible.
}
\end{deluxetable}

The $v \sin i$ measurements give lower limits on the true rotational velocities and may so be used to constrain the spin-axis inclinations of our targets. We assume that these brown dwarfs rotate as rigid spheres so that the equatorial rotation velocity is $v = 2 \pi R / P$, where $P$ is the photometric rotation period, and $R$ is the radius. We estimate the radii, masses, and ages by comparing our findings for surface gravities and effective temperatures to the $\log g$-$T_{\rm eff}$ grid in the evolutionary models of SM08.  Oblateness due to the rapid rotation (see Section~\ref{sec:breakup}; notes in Table~\ref{table:inclinations}) and the corresponding increase in equatorial radius produce a second-order effect, which we have ignored in these calculations.
Combining the radii ($R$), the photometric periods ($P_{\rm rot}$), and the spectroscopically determined projected rotational velocities ($v \sin i$), we calculate the inclinations ($i$) and the equatorial rotation velocities ($v_{\rm eq}$) of our targets (Table~\ref{table:inclinations}).  All three L and T dwarfs have equatorial velocities~$\gtrsim$100~km~s$^{-1}$, and 2MASS~J0348--6022 (T7) is seen near equator-on.  

All three objects are also likely substellar. At a spectral type of L3.5, 2MASS~J0407+1546 is the warmest and potentially most massive among our three L and T dwarfs. Its evolutionary model-dependent mass estimate is 0.037--0.073~$M_\odot$ (Table~\ref{table:inclinations}). Optical spectroscopy from \citet{Reid2008} does not reveal lithium absorption, so it must be $>0.060 M_\odot$ \citep[e.g.,][]{burrows_etal97}. This still leaves the estimated 0.060--0.073~$M_\odot$ mass of 2MASS~J0407+1546 mostly in the substellar ($<0.072 M_\odot$) domain.

The L3.5 dwarf 2MASS~J0407+1546 is also known to be chromospherically active based on the strong (60~\AA~equivalent width) H$\alpha$ emission reported by \citet{Reid2008}.  Its rapid rotation and H$\alpha$ emission may well indicate the presence of an aurora.  Based on radio detections of three L and T dwarfs with short (1.5~h--2.2~h) rotation periods, \citet{Kao2018} conclude that rapid rotation is key to powering auroral emissions via the electron cyclotron maser instability \citep{Hallinan2007, Hallinan2015}.  \citet{Kao2016, Pineda2017} and \citet{Richey-Yowell2020} further demonstrate that brown dwarf H$\alpha$ and radio luminosities and radio aurorae are correlated. It is possible that all three of our rapidly rotating brown dwarfs have strong dipole fields that power auroral emission \citep{Kao2018}. In particular, the near equator-on view of the T7 dwarf 2MASS~J0348--6022 makes it an excellent candidate for seeking pulses of circularly polarized electron cyclotron maser emission. This is already known from other rapidly rotating ultra-cool dwarfs seen at their equators \citep{Berger2001,Hallinan2007}.

\subsection{Proximity to Rotational Break-up and Oblateness}
\label{sec:breakup}

An upper limit on the spin rate of brown dwarfs exists from simple arguments of rotational stability. \citet{Konopacky2012} estimate that their two most rapidly rotating ultra-cool dwarfs, HD~130948C ($v\sin i=86$~km\,s$^{-1}$) and LP~349--45B ($v\sin i=83$~km\,s$^{-1}$) rotate at approximately 30\% of break-up speed.  The break-up periods for typical $>$1~Gyr-aged field brown dwarfs are in the tens of minutes. For example, a massive 0.07~$M_\odot$, 0.09~$R_\odot$ brown dwarf has $P_{\rm breakup} = 2\pi(R^3/GM)^{1/2} = 17$~min, while a low-mass 0.02~$M_\odot$, 0.12~$R_\odot$ brown dwarf has  $P_{\rm breakup} = 49$~min. These are approximately consistent with extrapolations from the shortest known (5~h) brown dwarf rotation period at 5 Myr \citep{Scholz_etal2015}, assuming that the fastest rotators at ~5 Myr remain the fastest when they contract and age. Using the evolutionary models of (non-accreting) brown dwarfs from \citet{Baraffe_etal2015}, by 3~Gyr conservation of angular momentum, dictates periods in the 10--70~min range \citep[e.g.,][]{Schneider_etal2018}. 

The 65--74~min periods of our three fast rotators are at the long end of this range. They would be near break-up only if they all had very low masses and large radii, i.e., were young brown dwarfs with low surface gravities. This is highly unlikely, given the wide range in spectral types (L3.5--T7) of our three rapid rotators, and the fact that their moderate-to-high surface gravities ($\log g\gtrsim 5.0$; Table~\ref{table:inclinations}) point to $>$0.1~Gyr ages. Using our measured RVs and precise proper motions and parallaxes determined from the Hawaii Infrared Parallax Program \citep{Liu2016, Best2020} or from Spitzer \citep{Kirkpatrick2019}, the BANYAN $\Sigma$ young moving group tool \citep{Gagne2018} reports that the space motions of all three L and T dwarfs are $\geq$99\% consistent with field-dwarf kinematics.  Only for 2MASS~J0407+1546 (L3.5) is there a 1\% chance of membership in the 40--50~Myr Argus association \citep{zuckerman19}, and \citet{Gagne2015} independently discuss that this object may either be $\sim$200~Myr old or have peculiar metallicity, based on weaker FeH and slightly weaker alkali line widths.  Thus, 2MASS~J0407+1546 may indeed be moderately young, even if it is not a member of any of the known young stellar moving groups.

To assess the proximity to break-up spin-velocity, we consider the effect of the centrifugal acceleration on surface gravity. Rapid rotation decreases the surface gravity near the equator, and may make the object appear younger. We can determine the surface gravity decrement due to the centrifugal acceleration using the inferred radii and equatorial velocities (Table~\ref{table:inclinations}). For our potentially fastest and largest rotator, the L8 dwarf 2MASS J1219+3128, the centrifugal acceleration is $a_c = v^2 / R = 1.6 \times 10^4$ cm s$^{-2}$ ($\log{a_c} = 4.2$), where $v_{\rm eq}=107$ km\,s$^{-1}$ and $R = 0.100 R_{\odot}$. The centrifugal acceleration thus reduces the surface gravity at the equator by about 13\%, when compared to the $\log g = 5.1 \pm 0.5$ surface gravity inferred from the photospheric model fitting (Table~\ref{table:inclinations}). While this does indicate that the rotation speed amounts to a significant fraction (35\%) of the break-up speed, we note that it has a minor effect on our ability to assess the surface gravity spectroscopically.

The rotational stability limit for brown dwarfs may not necessarily be set by the centrifugal levitation argument above.  The stability limit for the oblateness $f$ (fractional difference between polar and equatorial radii) of axisymmetric rotating polytropes for brown dwarf-like structures ($n\sim 1 $ to 1.5) is about 0.4 \citep{James1964}.  The Darwin-Radau relationship (e.g., \citealt{Barnes2003}) connects the oblateness, mass, radius, rotation, and moment of inertia for objects with smoothly varying interiors. Using the central values from Table~\ref{table:inclinations}, we compute oblateness factors of 0.08, 0.08 and 0.05 for the most (2MASS~J0348--6022, 2MASS~J1219+3128) and least (2MASS~J0407$+$1546) oblate objects.  This places the spin rates of both 2MASS~J0348--6022 and 2MASS~J1219+3128 at about 45\% of their rotational stability limits: closer to instability than indicated by the rigid-body rotation break-up velocity estimates. For comparison, Saturn, the most oblate planet in the solar system, has an oblateness of 0.1. The brown dwarfs have surface gravities about 100 times greater than Saturn but rotation rates 10 times faster. Since oblateness scales as $\Omega^2/g$ (where $\Omega$ is the rotation rate), it is not surprising the oblateness of these objects are comparable to that of Saturn.

Finally, the preceding discussion ignores the effect of any magnetic dynamo from the metallic hydrogen interior, which may be an important contributor to the energy balance in such rapid rotators, and may further limit the maximum spin velocity. So these three objects may be even closer to instability than indicated by estimates that ignore magnetic fields.

From an observational standpoint, the three rapid rotators delineate a clear lower boundary to the envelope of all 78 L-, T-, and Y-dwarf rotation periods measured to date (Fig.~\ref{fig:periodicity}). This limit holds over a broad range of spectral types, for objects that presumably have different ages.  Hence, $\sim$1~h may be close to a physical lower limit to the spin period of field-aged Jupiter-sized brown dwarfs.

Because of their significant oblateness, the three rapid rotators are potentially good targets for searches for polarized thermal emission (e.g., \citealt{Marley2011, deKok2011, Stolker2017}). Several surveys have been successful in detecting polarized thermal emission from brown dwarfs (e.g., \citealt{Menard2002, Zapatero2005, Miles2013, Miles2017a, Millar-Blanchaer2020}), which could be attributed to inhomogeneous cloud cover or oblateness.  Intriguingly, \citet{Miles2013} find that ultra-cool dwarfs with the fastest rotation ($v \sin i \geq$~60~km~s$^{-1}$) are more likely to exhibit linear polarization and at a larger degree than slower rotators.

\section{Conclusions}
\label{Conclusions}

We present a T7, L3.5, and an L8 dwarf with the shortest photometric periodicities measured to date: 1.08 h, 1.14 h, and 1.23 h. We confirm these extremely short rotation periods with moderate-dispersion spectroscopy and comparisons to Doppler-broadened model photospheres. The inferred $v \sin i$ value of the T7 dwarf 2MASS~J0348--6022 is the highest known to date for an ultra-cool dwarf. Combining the projected rotation velocities of our targets with their photometric periods and photospheric model-dependent radii, we determine their equatorial velocities. All three L and T dwarfs spin at $\gtrsim$100~km~s$^{-1}$ at their equators, and are the most rapidly spinning field ultra-cool dwarfs known to date.  As such, they are excellent candidates for seeking auroral radio emission, which has been linked to rapid rotation in ultra-cool dwarfs. We consider the role of the centrifugal acceleration on surface gravity, and find that, while the effect can be significant, at $\lesssim$0.1~dex in surface gravity it can be difficult to discern with current photospheric models. We find that the objects have oblateness factors of between 5\% and 8\%, which ranks them among the best targets for seeking net optical or infrared polarization. Given that the three rapid rotators presented in this paper appear to lie near a short-period limit of approximately 1~h across all brown dwarf spectral types, we consider it unlikely that rotation periods much shorter than 1~h exist for brown dwarfs.

\acknowledgments
{
We would like to thank the anonymous referee for their considerate and constructive comments that helped us improve this paper. 
Support for this work was provided by NASA through an award issued by JPL/Caltech (RSA \#1533692), by an NSERC Discovery Grant and the NSERC Canada Research Chairs program, by the Canada Space Agency (grant \#18FAWESC13), and by an Ontario Graduate Scholarship.
This work is based in part on observations made with the Spitzer Space Telescope, which is operated by the Jet Propulsion Laboratory, California Institute of Technology under a contract with NASA. 
The authors wish to recognize and acknowledge the very significant cultural role and reverence that the summit of Maunakea has always had within the indigenous Hawaiian community. We are most fortunate to have the opportunity to conduct observations from this mountain. This paper includes data gathered with the 6.5 m Magellan Telescopes located at Las Campanas Observatory, Chile. 
IRAF is distributed by the National Optical Astronomy Observatory, which is operated by the Association of Universities for Research in Astronomy (AURA) under a cooperative agreement with the National Science Foundation.
}

\facilities{Spitzer (IRAC), Magellan:Baade (FIRE), Gemini:Gillett (GNIRS)}

\software{PyRAF \citep{pyraf}, FIREHOSE v2 \citep{Gagnezenodo}, BANYAN $\Sigma$ \citep{Gagne2018}}

\appendix

The full list of rotation periods shown in Figure 13 is given in Table A1 with references.

\setcounter{table}{0}
\renewcommand{\thetable}{A\arabic{table}}

\startlongtable
\begin{deluxetable}{l l l c c c c} 
\tabletypesize{\scriptsize}
\tablecolumns{7} 
\tablewidth{0pt}
\tablecaption{Known L-, T-, and Y-Dwarf Rotation Periods \label{table:allperiods} }
\tablehead{
\colhead{Object} & \colhead{RA} & \colhead{DEC} & \colhead{Spectral} & \colhead{Period} & \colhead{Period} & \colhead{Reference}\\[-0.05cm]
\colhead{} & \colhead{} & \colhead{} & \colhead{Type} & \colhead{(h)} & \colhead{Uncertainty (h)} & \colhead{}
}
\startdata 
2MASS J00132229-1143006	&	00 13 22.2	&	$-$11 43 00.6	&	T3	&	2.8	&	\nodata	&	(1)	\\
LSPM J0036+1821	&	00 36 16.1	&	+18 21 10.2	&	L3.5	&	2.7	&	0.3	&	(2), (3), (4), (5)	\\
2MASS J00452143+1634446	&	00 45 21.4	&	+16 34 44.7	&	L2$\beta$	&	2.4	&	0.1	&	(6), (7)	\\
2MASS J00470038+6803543	&	00 47 00.3	&	+68 03 54.3	&	L7 	&	16.4	&	0.2	&	(8), (9)	\\
2MASS J00501994-3322402	&	00 50 19.9	&	$-$33 22 40.2	&	T7	&	1.55	&	0.02	&	(2)	\\
2MASSI J0103320+193536	&	01 03 32.0	&	+19 35 36.1	&	L6	&	2.7	&	0.1	&	(2)	\\
2MASS J01075242+0041563	&	01 07 52.4	&	+00 41 56.3	&	L8	&	5	&	\nodata	&	(2)	\\
GU Psc B	&	01 12 36.5	&	+17 04 29.9	&	T3.5	&	5.9	&	0.7	&	(10), (11)	\\
2MASS 0122-2439 b	&	01 22 50.8	&	$-$24 39 51.6	&	L5	&	6	&	\nodata	&	(12)	\\
SIMP J013656.5+093347.3	&	01 36 56.6	&	+09 33 47.3	&	T2.5	&	2.3895	&	0.0005	&	(13), (14), (5), (15), (16), (17), (1)	\\
2MASS J01383648-0322181	&	01 38 36.4	&	$-$03 22 18.1	&	T3	&	3.2	&	\nodata	&	(1)	\\
DENIS J025503.3-470049	&	02 55 03.6	&	$-$47 00 51.3	&	L8	&	7.4	&	\nodata	&	(18), (19), (20)	\\
2MASS J03480772-6022270	&	03 48 07.7	&	$-$60 22 27.0	&	T7	&	1.08	&	0.005	&	(21), (22)	\\
2MASS J04070752+1546457 	&	04 07 07.5	&	+15 46 45.5	&	L3.5	&	1.23	&	0.01	&	(21)	\\
2MASSI J0423485-041403 	&	04 23 48.5	&	$-$04 14 03.2	&	L7	&	1.47	&	0.13	&	(17), (23)	\\
2MASS J05012406-0010452	&	05 01 24.0	&	$-$00 10 45.5	&	L3	&	15.7	&	0.2	&	(6), (7)	\\
Beta Pic b	&	05 47 17.0	&	$-$51 03 59.4	&	L1	&	8.1	&	1.0	&	(24)	\\
2MASS J05591914-1404488	&	05 59 19.1	&	$-$14 04 49.2	&	T4.5	&	10	&	\nodata	&	(14)	\\
AB Pic B	&	06 19 12.9	&	$-$58 03 20.9	&	L1	&	2.12	&	\nodata	&	(12)	\\
2MASS J07003664+3157266	&	07 00 36.7	&	+31 57 25.5	&	L3.5	&	3.79	&	1.3	&	(25)	\\
2MASS J07464256+2000321A	&	07 46 42.4	&	+20 00 32.6	&	L0.5	&	3.32	&	0.15	&	(4)	\\
2MASS J07584037+3247245	&	07 58 40.3	&	+32 47 24.5	&	T2 	&	4.9	&	0.2	&	(14)	\\
2MASS J08173001-6155158	&	08 17 29.9	&	$-$61 55 15.6	&	T6.5 	&	2.8	&	0.2	&	(14)	\\
2MASSI J0825196+211552	&	08 25 19.6	&	+21 15 51.5	&	L7.5	&	7.6	&	\nodata	&	(2), (26)	\\
2MASS J08283419-1309198	&	08 28 34.1	&	$-$13 09 19.8	&	L2	&	2.9	&	\nodata	&	(27)	\\
2MASS J08354256-0819237	&	08 35 42.5	&	$-$08 19 23.3	&	L4.5	&	3.1	&	\nodata	&	(27), (22)	\\
LP 261-75 B	&	09 51 05.4	&	+35 58 02.1	&	L6	&	4.78	&	0.95	&	(28)	\\
2MASSI J1043075+222523	&	10 43 07.5	&	+22 25 23.6	&	L8	&	2.21	&	0.14	&	(17)	\\
2MASS J10433508+1213149	&	10 43 35.0	&	+12 13 14.9	&	L9	&	3.8	&	0.2	&	(2)	\\
2MASSW J1047539+212423	&	10 47 53.8	&	+21 24 23.4	&	T6.5	&	1.741	&	0.007	&	(29), (30), (17)	\\
Luhman 16A	&	10 49 19.0	&	$-$53 19 10	&	L7.5	&	6.94	&	\nodata	&	(31), (32), (33), (34), (35)	\\
Luhman 16B	&	10 49 18.9	&	$-$53 19 09	&	T0.5	&	5.28	&	\nodata	&	(31), (32), (36), (33), (34), (35)	\\
2MASS J10521350+4422559	&	10 52 13.5	&	+44 22 55.9	&	T0.5	&	3	&	\nodata	&	(37)	\\
DENIS J1058.7-1548	&	10 58 47.8	&	$-$15 48 17.2	&	L3	&	4.1	&	0.2	&	(2), (38)	\\
2MASS J11193254-1137466	&	11 19 32.5	&	$-$11 37 46.6	&	L7	&	3.02	&	0.04	&	(39)	\\
2MASS J11225550+2550250	&	11 22 55.5	&	+25 50 25.0	&	T6	&	1.93	&	0.12	&	(40), (41)	\\
DENIS J112639.9-500355	&	11 26 39.8	&	$-$50 03 54.8	&	L4.5	&	3.2	&	0.3	&	(2), (14)	\\
WISEA J114724.10-204021.3	&	11 47 24.2	&	$-$20 40 20.4	&	L7	&	19.39	&	0.33	&	(39)	\\
2MASSW J1207334-393254 b	&	12 07 33.5	&	$-$39 32 54.4	&	L5	&	10.7	&	1.2	&	(42), (12)	\\
HD 106906 B	&	12 17 52.6	&	$-$55 58 26.6	&	L2.5	&	4	&	\nodata	&	(43)	\\
2MASS J12195156+3128497	&	12 19 51.5	&	+31 28 49.7	&	L8	&	1.14	&	0.03	&	(21), (26)	\\
2MASS J12373919+6526148	&	12 37 39.1	&	+65 26 14.8	&	T6.5	&	2.28	&	0.1	&	(17)	\\
VHS J1256-1257B	&	12 56 01.8	&	$-$12 57 27.6	&	L7	&	22.5	&	0.4	&	(44)	\\
Ross 458 C	&	13 00 42.0	&	+12 21 15.0	&	T8	&	6.75	&	1.58	&	(45)	\\
Kelu-1	&	13 05 40.1	&	$-$25 41 05.8	&	L2	&	1.8	&	\nodata	&	(46), (47)	\\
2MASS J13243553+6358281	&	13 24 35.5	&	+63 58 28.1	&	T2	&	13.2	&	\nodata	&	(16), (2)	\\
WISE J140518.39+553421.3	&	14 05 18.3	&	+55 34 21.3	&	Y0	&	8.5	&	\nodata	&	(48)	\\
2MASS J14252798-3650229	&	14 25 27.9	&	$-$36 50 23.2	&	L5	&	3.7	&	0.8	&	(14), (7), (6)	\\
DENIS J145407.8-660447	&	14 54 07.9	&	$-$66 04 47.4	&	L3.5	&	2.57	&	0.002	&	(47)	\\
2MASSW J1507476-162738	&	15 07 47.6	&	$-$16 27 40.1	&	L5	&	2.5	&	0.1	&	(2), (15)	\\
SDSS J151114.65+060742.9	&	15 11 14.6	&	+06 07 43.1	&	T2	&	11	&	2	&	(2)	\\
2MASS J15164306+3053443	&	15 16 43.0	&	+30 53 44.3	&	T0.5	&	6.7	&	\nodata	&	(2)	\\
2MASS J15394189-0520428	&	15 39 41.9	&	$-$05 20 42.7	&	L3.5	&	2.51	&	1.6	&	(25), (49)	\\
2MASS J16154255+4953211	&	16 15 42.5	&	+49 53 21.1	&	L4$\beta$	&	24	&	\nodata	&	(2)	\\
2MASS J16291840+0335371	&	16 29 18.4	&	+03 35 37.1	&	T2	&	6.9	&	2.4	&	(14)	\\
2MASS J16322911+1904407	&	16 32 29.1	&	+19 04 40.7	&	L8	&	3.9	&	0.2	&	(2)	\\
2MASSI J1721039+334415	&	17 21 03.6	&	+33 44 16.9	&	L3	&	2.6	&	0.1	&	(2)	\\
JWISE J173835.53+273259.0	&	17 38 35.5	&	+27 32 59.0	&	Y0	&	6	&	0.1	&	(50)	\\
2MASS J17502385+4222373	&	17 50 23.8	&	+42 22 37.3	&	T2	&	2.7	&	0.2	&	(14)	\\
2MASS J17534518-6559559	&	17 53 45.1	&	$-$65 59 55.6	&	L4	&	$\geq$50	&	\nodata	&	(2)	\\
2MASS J18071593+5015316	&	18 07 15.9	&	+50 15 31.6	&	L1.5	&	1.71	&	0.3	&	(25)	\\
2MASS J18212815+1414010	&	18 21 28.1	&	+14 14 00.8	&	L4.5	&	4.2	&	0.1	&	(2), (15)	\\
2MASS J18283572-4849046	&	18 28 35.7	&	$-$48 49 04.6	&	T5.5	&	5	&	0.6	&	(14)	\\
2MASS J19064801+4011089	&	19 06 48.0	&	+40 11 08.5	&	L1	&	8.9	&	\nodata	&	(51)	\\
2MASSI J2002507-052152	&	20 02 50.7	&	$-$05 21 52.5	&	L5.5	&	8	&	2	&	(7)	\\
2MASS J20360316+1051295	&	20 36 03.1	&	+10 51 29.5	&	L3	&	1.45	&	0.55	&	(25)	\\
PSO J318.5338-22.8603	&	21 14 08.0	&	$-$22 51 35.8	&	L7	&	8.45	&	0.05	&	(7), (52), (53), (54)	\\
HD 203030B	&	21 18 58.9	&	+26 13 46.1	&	L7.5	&	7.5	&	0.6	&	(55)	\\
2MASS J21392676+0220226	&	21 39 26.7	&	+02 20 22.6	&	T2	&	7.614	&	0.178	&	(15), (56), (14), (16)	\\
HN Peg B	&	21 44 28.4	&	+14 46 07.7	&	T2.5	&	15.4	&	0.5	&	(57), (2)	\\
2MASSW J2148162+400359	&	21 48 16.2	&	+40 03 59.3	&	L6	&	19	&	4	&	(2)	\\
2MASS J21483578+2239427	&	21 48 35.7	&	+22 39 42.7	&	T1	&	2.4	&	0.4	&	(1)	\\
2MASSW J2208136+292121	&	22 08 13.6	&	+29 21 21.5	&	L3$\gamma$	&	3.5	&	0.2	&	(2)	\\
2MASS J22153705+2110554	&	22 15 37.0	&	+21 10 55.4	&	T1	&	5.2	&	0.5	&	(1)	\\
2MASS J22282889-4310262	&	22 28 28.8	&	$-$43 10 26.2	&	T6	&	1.41	&	0.01	&	(2), (23), (58), (14), (15)	\\
2MASS J22393718+1617127	&	22 39 37.1	&	+16 17 12.7	&	T3	&	3.4	&	\nodata	&	(1)	\\
2MASS J22443167+2043433	&	22 44 31.6	&	+20 43 43.3	&	L6	&	11	&	2	&	(8), (18), (7)	\\
2MASS J23312378-4718274	&	23 31 23.7	&	$-$47 18 27.4	&	T5.5	&	2.9	&	0.9	&	(23)	\\
\enddata 
\tablecomments{
Where multiple references are given, we have adopted the spectral type and period value from the first reference. Additional L3--T8 periods compiled in \citet{Crossfield2014} that have not withstood independent confirmation so we do not include them here.
} 
\tablerefs{
~(1)~\citet{Eriksson2019};
~(2)~\citet{Metchev2015};
~(3)~\citet{Berger2005};
~(4)~\citet{Harding2013};
~(5)~\citet{Croll2016};
~(6)~\citet{Vos2020};
~(7)~\citet{Vos2019};
~(8)~\citet{Vos2018};
~(9)~\citet{Lew2016};
~(10)~\citet{Naud2017};
~(11)~\citet{Lew2020};
~(12)~\citet{Zhou2019};
~(13)~\citet{Artigau2009};
~(14)~\citet{Radigan2014a};
~(15)~\citet{Yang2016};
~(16)~\citet{Apai2017};
~(17)~\citet{Kao2018};
~(18)~\citet{Morales2006};
~(19)~\citet{Koen2005a};
~(20)~\citet{Koen2005b};
~(21)~This work;
~(22)~\citet{Wilson2014};
~(23)~\citet{Clarke2008};
~(24)~\citet{Snellen2014};
~(25)~\citet{Miles2017c};
~(26)~\citet{Buenzli2014};
~(27)~\citet{Koen2004};
~(28)~\citet{Manjavacas2018};
~(29)~\citet{Allers2020};
~(30)~\citet{Williams2015};
~(31)~\citet{Apai2021};
~(32)~\citet{Biller2013};
~(33)~ \citet{Buenzli2015a};
~(34)~\citet{Buenzli2015b};
~(35)~\citet{Karalidi2016};
~(36)~\citet{Gillon2013};
~(37)~\citet{Girardin2013};
~(38)~\citet{Heinze2013};
~(39)~\citet{Schneider2018};
~(40)~\citet{Williams2017};
~(41)~\citet{Route2016};
~(42)~\citet{Zhou2016};
~(43)~\citet{Zhou2020};
~(44)~\citet{Bowler2020};
~(45)~\citet{Manjavacas2019};
~(46)~\citet{Clarke2002};
~(47)~\citet{Koen2013};
~(48)~\citet{Cushing2016};
~(49)~\citet{Koen2013a};
~(50)~\citet{leggett2016};
~(51)~\citet{Gizis2015};
~(52)~\citet{Biller2015};
~(53)~ \citet{allers_etal16};
~(54)~\citet{Biller2018};
~(55)~\citet{Miles2019};
~(56)~\citet{Radigan2012};
~(57)~\citet{Zhou2018};
~(58)~\citet{Buenzli2012}.
}
\end{deluxetable}

\bibliography{main}

\end{document}